%% file: main.tex
\documentclass[conference, romanappendices]{IEEEtran}
\IEEEoverridecommandlockouts

\usepackage{cite}
\usepackage{amsmath,amssymb,amsfonts}
\usepackage{algorithmic}
\usepackage{graphicx}
\usepackage{textcomp}
\def\BibTeX{{\rm B\kern-.05em{\sc i\kern-.025em b}\kern-.08em
    T\kern-.1667em\lower.7ex\hbox{E}\kern-.125emX}}
\usepackage{cite}
\usepackage{amsmath,amssymb,amsfonts}
\usepackage{algorithmic}
\usepackage{graphicx}
\usepackage{textcomp}
\usepackage[dvipsnames,svgnames]{xcolor}
\usepackage{fancyhdr}
\usepackage[hyphens]{url}
\usepackage[bookmarks=true,breaklinks=true,letterpaper=true,colorlinks,citecolor=blue,linkcolor=blue,urlcolor=blue]{hyperref}

\usepackage{sourcecodepro}

\usepackage{amsmath,amsfonts}
\usepackage{amsthm}
\usepackage{graphicx}
\usepackage{textcomp}
\usepackage{colortbl} 
\usepackage{hhline}
\usepackage{tcolorbox}
\usepackage{ragged2e}
\usepackage{marvosym}
\usepackage{enumitem}
\theoremstyle{definition}
\newtheorem{definition}{Definition}[section]
\usepackage{algorithm}
\usepackage{pifont}

\newcommand{\synthlc}{\textsc{SynthLC}}

\newcommand{\chkm}{$\checkmark$}
\newcommand{\ct}{CT}

\newcommand{\upaths}{\textsc{$\mu$path}s}
\newcommand{\upath}{\textsc{$\mu$path}}
\newcommand{\rtltouspec}{\textsc{rtl2$\mu$spec}}

\newcommand{\uhb}{\textsc{$\mu$hb}}

\newcommand{\Rupath}{$\mathnormal{R}_{\mu \textsc{path}}$}

\newcommand{\uspec}{\textsc{$\mu$spec}}

\newcommand{\ufsms}{$\mu$FSMs}
\newcommand{\ufsm}{$\mu$FSM}
\newcommand{\lbbar}{\{\kern-0.5ex|}
\newcommand{\rbbar}{|\kern-0.5ex\}}
\newcommand{\lbkt}{[\kern-0.3ex[}
\newcommand{\rbkt}{]\kern-0.3ex]}

\newcommand{\llt}{$l$}

\newcommand{\PathFinder}{\textsc{rtl2m$\mu$path}}

\newcommand{\iparg}[1]{\textbf{#1}}
\newcommand{\itarg}[1]{\textbf{#1}}

\usepackage[normalem]{ulem}

\usepackage{circledsteps}
\usepackage{tikz}
\usepackage{array}
\usepackage{hhline}
\usepackage{multirow}
\usepackage{caption}
\usepackage{subcaption}
\usepackage{listings}
\lstdefinestyle{psuedocode}{
        linewidth=180mm,
        frame=none,
        basicstyle=\fontfamily{SourceCodePro-TLF}\selectfont\scriptsize, %
        identifierstyle=\color{black},
        commentstyle=\color{black},
        keywordstyle=\color{blue},
        keywords={assume, cover, s_eventually, and, assert, always, begin, end, } 
        escapechar={^},
        belowskip=2mm,
}

\lstdefinestyle{uspecStyle}{
    belowcaptionskip=1\baselineskip,
    breaklines=true,
    frame=none,
    numbers=none,
    mathescape=true,
    basicstyle=\fontfamily{SourceCodePro-TLF}\selectfont\scriptsize, 
    keywordstyle=[1]\bfseries\color{purple},
    keywordstyle=[2]\bfseries\color{blue!50!black},
    keywordstyle=[3]\bfseries\color{red!50!black},
    keywordstyle=[4]\bfseries\color{black},
    commentstyle=\itshape\color{green!40!black},
    comment = [l]{\%},
    keywords=[1]{forall},
    keywords=[2]{microops},
    keywords=[3]{IsMultiply, IsAnyRead, AddEdges, ProgramOrder, AddEdge, EdgeExists, IsAnyWrite, SamePA, SameData, NoWritesInBetween},
    keywords=[4]{Axiom},
    escapechar={^},
}

\newcommand{\svaarg}[1]{\textcolor{blue}{#1}}
\newcommand{\svaterm}[1]{{\fontfamily{SourceCodePro-TLF}\selectfont\small\bfseries\color{red!40!black} #1}}
\newcommand{\myttt}[1]{{\fontfamily{SourceCodePro-TLF}\selectfont\small #1}}

\newcommand{\svaplus}[1]{\color{green!40!black}{\textbf{#1}}}

\lstdefinestyle{svaListing2}{
    language=verilog,
    basicstyle=\fontfamily{SourceCodePro-TLF}\selectfont\scriptsize,
    commentstyle=\color{green!40!black}, 
    breaklines=true,frame=none,tabsize=2,escapechar={*},
    keywords={assume, assert, cover, always, s_eventually, and, module, input, logic, output, enum, always_ff, posedge, or negedge, begin},
    keywordstyle=\bfseries\color{red!40!black}
}

\lstdefinestyle{svaListing}{
    language=verilog,
    basicstyle=\fontfamily{SourceCodePro-TLF}\selectfont\scriptsize,
    commentstyle=\color{green!40!black}, 
    breaklines=true,frame=none,tabsize=2,escapechar={^},
    keywords={assume, assert, cover, always, s_eventually, and, module, input, logic, output, enum, always_ff, posedge, or negedge, begin},
    keywordstyle=\bfseries\color{red!40!black}
}

\lstdefinestyle{fooListing}{
    language=verilog,
    basicstyle=\fontfamily{SourceCodePro-TLF}\selectfont\scriptsize,
    commentstyle=\color{green!40!black}, 
    xleftmargin=.05in,
    breaklines=true,frame=none,tabsize=1,escapechar={^},
    keywords={assume, assert, cover, always, s_eventually, and, module, input, logic, output, enum, always_ff, posedge, or negedge, begin},
    keywordstyle=\bfseries\color{red!40!black}
}

\definecolor{mygray}{cmyk}{0,0,0,0.27}
\definecolor{myyellow}{HTML}{F7F056} 
\definecolor{myred}{HTML}{DC050C}
\definecolor{myblue}{HTML}{7BAFDE} 
\definecolor{mygray}{HTML}{ADADAD}

\definecolor{myorange}{HTML}{F1932D} 
\definecolor{lighyellow}{HTML}{fffdaf}
\definecolor{lightorange}{HTML}{ffa88b} 
\newcommand{\myhlt}[2]{{\setlength{\fboxsep}{0.5pt}\colorbox{#1}{#2\vphantom{Ay}}}}

\newcommand{\hlt}[1]{{\setlength{\fboxsep}{0pt}\colorbox{lighyellow}{#1}}}
\newcommand{\hltim}[1]{{\setlength{\fboxsep}{0pt}\colorbox{lightorange}{#1}}}
\newcommand{\inst}[1]{$\mathtt{#1}$}

\usepackage{ctable}
\newcommand{\M}{$\mathnormal{M}$}
\newcommand{\R}{$\mathnormal{R}$}

\renewcommand{\theequation}{\thesection.\arabic{equation}}

\IEEEaftertitletext{\vspace{-1\baselineskip}}

\begin{document}

\title{\textsc{rtl2m$\mu$path}: Multi-\upath{} Synthesis with Applications to Hardware Security Verification}
\IEEEoverridecommandlockouts
\makeatletter
\newcommand{\linebreakand}{%
  \end{@IEEEauthorhalign}
  \hfill\mbox{}\par
  \mbox{}\hfill\begin{@IEEEauthorhalign}
}
\makeatother
\author{\IEEEauthorblockN{Yao Hsiao}
\IEEEauthorblockA{\textit{Stanford University} \\ 
yaohsiao@stanford.edu}
\and
\IEEEauthorblockN{Nikos Nikoleris}
\IEEEauthorblockA{\textit{Arm} \\
Nikos.Nikoleris@arm.com}
\and
\IEEEauthorblockN{Artem Khyzha}
\IEEEauthorblockA{\textit{Arm} \\
Artem.Khyzha@arm.com}
\and
\IEEEauthorblockN{Dominic P. Mulligan*\thanks{*Work done while at Arm.}}
\IEEEauthorblockA{\textit{Amazon Web Services} \\
dominic.p.mulligan@gmail.com}
\linebreakand %
\IEEEauthorblockN{Gustavo Petri*}
\IEEEauthorblockA{\textit{Amazon Web Services} \\
gfpetri@amazon.co.uk}
\and
\IEEEauthorblockN{Christopher W. Fletcher}
\IEEEauthorblockA{\textit{University of California, Berkeley} \\
cwfletcher@berkeley.edu}
\and
\IEEEauthorblockN{Caroline Trippel}
\IEEEauthorblockA{\textit{Stanford University} \\
trippel@stanford.edu}

}

\maketitle

\input{00-abstract}
\input{01-intro-microfied}
\input{02-background}
\input{03-upath-formalism}
\input{03-tspec}
\input{04-tsynth-approach}
\input{05-tsynth-tool}
\input{06-case-study}

\input{07-results}
\input{08-related}
\input{09-conclusions}
\input{10-ack}
\input{todo}
\bibliographystyle{IEEEtranS}
\bibliography{references}
\clearpage 

\appendices
\input{11-ae}

\end{document}

%% file: 00-abstract.tex
\begin{abstract}
The \textit{Check} tools automate formal memory consistency model and security verification of processors by analyzing \textit{abstract models} of microarchitectures, called \uspec{} models.
Despite the efficacy 
of this approach, a verification gap between \uspec{} models, which must be manually written, and RTL limits the \textit{Check} tools' broad adoption.
Our prior work, called \rtltouspec{}, narrows this gap
by automatically synthesizing formally verified \uspec{} models from SystemVerilog implementations of simple processors. But, \rtltouspec{} assumes input designs where an instruction (e.g., a load) cannot exhibit more than one  \textit{microarchitectural execution path} (\upath{}, e.g., a cache hit or miss path)---its \textit{single-execution-path assumption}.

In this paper, we first propose an automated approach and tool, called \PathFinder{}, that resolves \rtltouspec{}'s single-execution-path assumption. Given a SystemVerilog processor design, instruction encodings, and modest design metadata, \PathFinder{} finds a complete set of formally verified \upaths{} for each instruction. Next, we make an important observation: an instruction that can exhibit more than one \upath{} strongly indicates the presence of a microarchitectural side channel in the input design. Based on this observation, we then propose an automated approach and tool, called \synthlc{}, that extends \PathFinder{} 
with a symbolic information flow analysis
to support synthesizing a variety of formally verified \emph{leakage contracts} from SystemVerilog processor designs. Leakage contracts are foundational to state-of-the-art defenses against hardware side-channel attacks. \synthlc{} is the \emph{first automated methodology} for formally verifying hardware adherence to them.

\end{abstract}

%% file: 01-intro-microfied.tex
\section{Introduction}
\label{sec:intro}

A common strategy to improve the
efficacy
of a formal verification procedure is to analyze an \textit{abstract model} of the target system, which omits irrelevant design details~\cite{mcmillan1993symbolic,clarke1994model,long1993model}. This approach is exemplified by the \textit{Check} tools~\cite{checkweb}, which automate formal memory consistency model~\cite{pipecheck, ccicheck, coatcheck, tricheck, rtlcheck, pipeproof} and security~\cite{checkmate} verification of processors.


At their core, the Check tools conduct \textit{microarchitectural happens-before} (\uhb{}) analysis~\cite{pipecheck}, which
models hardware-specific program executions as directed \uhb{} graphs (Fig.~\ref{fig:zero-skip-uhb}). 
A node in a \uhb{} graph represents a microarchitectural event, such as a dynamic program instruction (column label)
updating a particular set of hardware state elements (row label)~\cite{rtl2uspec:hsiao:2021}.
Directed edges denote happens-before relationships~\cite{lamport:happensbefore}. 

To facilitate \uhb{} analysis, the Check tools analyze a microarchitecture in the guise of an \textit{axiomatic \uspec{} model}---an abstract model of a microarchitecture, which omits irrelevant RTL 
details~\cite{coatcheck}. In particular, a \uspec{} model is a set of first-order logic axioms (rules) that describe how to construct \uhb{} graphs to model hardware-specific program executions.
Axioms encode (i) all \textit{microarchitectural execution paths} (\upaths{}---our term) for each implemented instruction, to instantiate column-wise nodes/edges (as in Fig.~\ref{fig:zero-skip-uhb}) and (ii) all possible \textit{microarchitectural dependencies} between pairs of executed instructions, to instantiate edges between columns~\cite{rtl2uspec:hsiao:2021}.



%

Despite finding bugs in real hardware~\cite{pipecheck, tricheck, c11counterexamples, checkmate, spectremeltdownprime}, the \textit{Check} tools have not achieved broad adoption due to a verification gap between \uspec{} models, which must be manually written, and RTL.

Our prior work narrows this gap via an automated approach and tool, called \rtltouspec{}, for synthesizing formally verified \uspec{} models from simple SystemVerilog processor designs~\cite{rtl2uspec:hsiao:2021}. 
However, \rtltouspec{} possesses a critical limitation: it cannot discover more than one 
\upath{} 
per implemented instruction.
Hence, designs where instructions may exhibit more than one \upath{} are not supported---the \textit{single-execution-path assumption}~\cite{rtl2uspec:hsiao:2021}.
This restriction is incompatible with pervasive hardware features like: caches, which create hit and miss paths for memory instructions; variable-time functional units (e.g., serial dividers), which create a few to many path possibilities for certain instructions;
speculation, which creates commit and squash paths for virtually all instructions; and multiple copies of the same functional unit, where certain instructions may take a path that uses any one of the units.
We further show that it precludes designs with microarchitectural side channels, making \rtltouspec{}-synthesized \uspec{} models useless for \textit{Check}-based hardware security verification~\cite{checkmate}.

\subsection{This Paper}
\label{sec:intro:thispaper}
This paper makes three key contributions.


\paragraph*{Foundation: Synthesizing Microarchitectural Execution Paths}
\textbf{Our first contribution} is an automated
approach and tool, called \PathFinder{}, that resolves \rtltouspec{}'s single-execution-path assumption.
Given a SystemVerilog processor design, instruction encodings, and modest (mostly standard) design metadata, \PathFinder{} uses static netlist analysis, linear temporal logic (LTL)~\cite{pnueli:ltl,manna:ltl-textbook} property generation (from property templates), and model checking~\cite{baier:principles_of_model_checking, clarke:model_checking} to find
a complete set of formally verified \upaths{} (\uhb{} graph columns) for each instruction.
Automated \upath{} synthesis with \PathFinder{}
is enabled by two key aspects of its design.

First, we extend the \uhb{} graph formalism from prior work with \textit{cycle-accurate} timing information. In this paper, a \uhb{} node represents an instruction updating a particular set of hardware state elements \textit{in a specific cycle}; edges encode \textit{one-cycle} happens-before relationships. 
Cycle-accurate \upaths{} enable \PathFinder{} to distinguish executions where the same set of state elements is updated
a different number of times.


Second, \PathFinder{} recognizes a \uhb{} node during an instruction's execution on a microarchitecture when the instruction
\textit{visits} (i.e., \textit{occupies})
a particular \textit{performing location} (PL) in a given cycle. Similar to a pipeline stage, but more granular, a PL represents a \textit{step}
of an instruction's execution during which it has exclusive write access to a particular subset of design states.
That is, PLs encapsulate instructions' state updates per cycle.
We show that PLs are precisely captured by certain finite state machines within a processor's control-path (\S\ref{sec:pls}).
By conceptualizing \uhb{} nodes as instructions' visits to PLs,
\PathFinder{} supports {speculative}, 
superscalar, and out-of-order pipelines, {plus caches}. 

\paragraph*{Observation: Security Implications of \upath{} Variability}
In designing \PathFinder{}, we make an important observation about
how programs leak their private data in hardware side-channel attacks.
Briefly, these attacks are often defined using a telecommunications analogy~\cite{dawg:kiriansky}:
a \textit{transmit instruction} (or \textit{transmitter}, in the victim program) modulates a \textit{channel} (hardware resource) in an operand-dependent manner, and a \textit{receiver} (attacker) observes the channel modulation to infer the operand value~\cite{dawg:kiriansky}. We observe that: \vspace{-5pt}
\begin{tcolorbox}[left=2pt,right=2pt,top=0pt,bottom=0pt]
\textbf{Observation I:} When a transmitter modulates a channel in an operand-dependent manner, it creates operand-dependent
\upath{} variability (${>}1$ \upath{}) for one or more \textit{transponder instructions} (or \textit{transponders}---our term).
A receiver observes a transmitter's distinct channel modulations as distinct \upaths{} for transponder(s).
\end{tcolorbox}
\vspace{-5pt}

\begin{figure}[!t]
    \centering
    \includegraphics[width=0.8\linewidth]{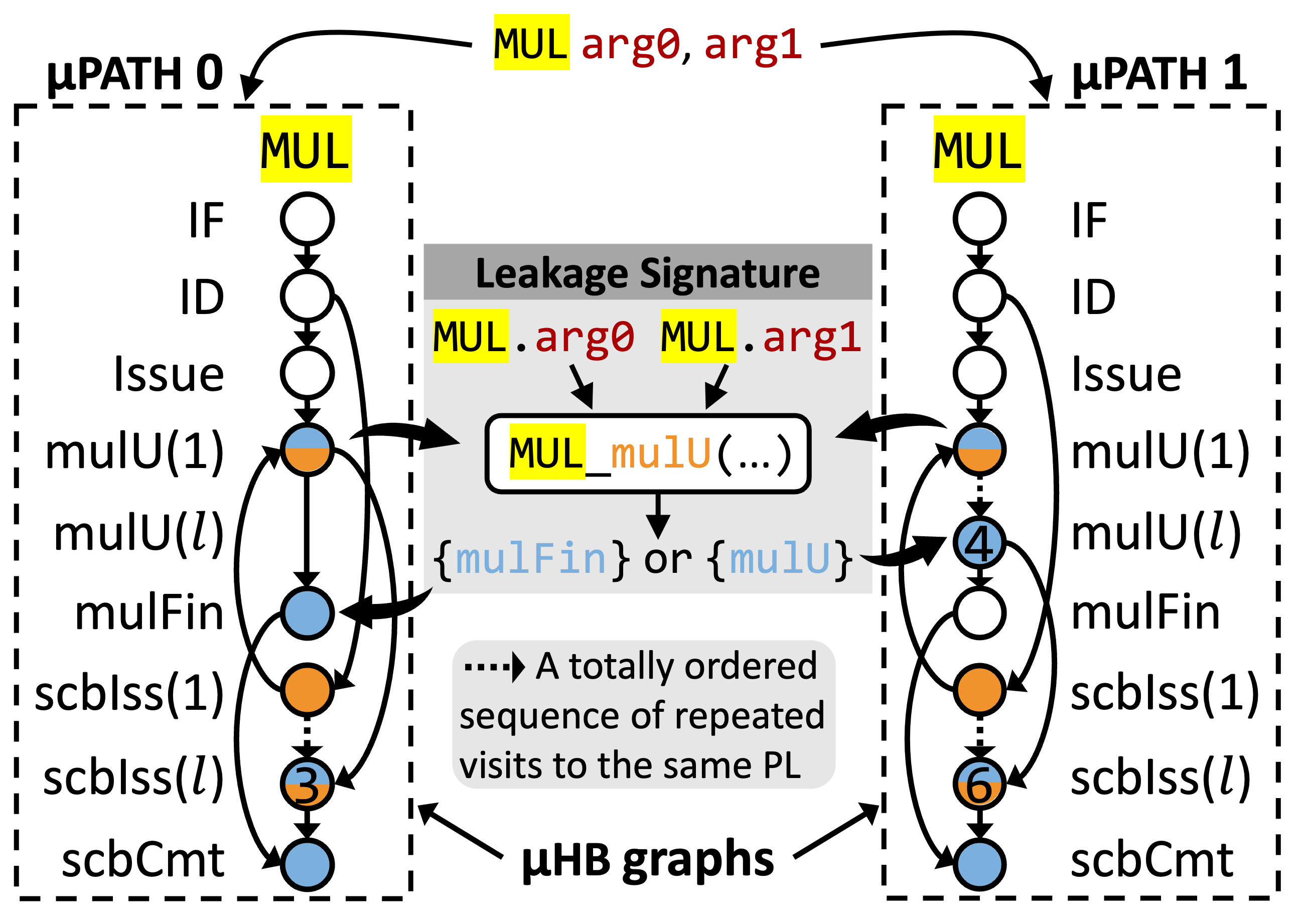}
    \caption{Two \upaths{} for \inst{MUL} on CVA6-MUL
    (\S\ref{sec:intro:thispaper}) and a leakage signature, which defines {\setlength{\fboxsep}{0pt}\colorbox{myyellow}{transponder}} \inst{MUL}'s \upath{} variability as a function of its own operands following its visit to the \inst{mulU} PL.
    Row({\inst{1}/\llt{}}): 1st/\llt{}-th visit to Row.
    Node label: value of \llt{} for the \upath{}. 
   }
    \label{fig:zero-skip-uhb}
\end{figure}
As an example, consider Fig.~\ref{fig:zero-skip-uhb}, which shows two \upaths{} for a 32-bit multiply (\inst{MUL}) instruction executing on CVA6-MUL, a variant of the RISC-V CVA6 CPU~\cite{cva6} that implements the \textit{zero-skip multiply} optimization~\cite{andrysco:subnormal,Mult_leaky, arm7TDMI}.
On this design, a \inst{MUL} will spend one cycle in the multiplication unit if it has at least one zero operand; else, it will spend four cycles~\cite{arm7TDMI}.
Such a \inst{MUL} is a transmitter~\cite{pandora:isca:21}: it occupies a hardware resource for an operand-dependent number of cycles.

Clearly, a \inst{MUL}
creates \upath{} variability for itself: it may visit $\mathtt{mulU}$, the \textit{multiplication unit} PL, for one ($\mu\textsc{path}~0$) or four ($\mu\textsc{path}~1$) consecutive cycles.
More subtly, a \inst{MUL} may
create multiple \upaths{} for subsequent (in program order), concurrently in-flight instructions, which may stall behind the \inst{MUL} for one to five cycles before committing (after completing).
So, \inst{MUL} transmitters implicate themselves \textit{and} younger, concurrent
instructions as transponders.
A receiver observes a \inst{MUL}'s distinct channel modulations as distinct
\upaths{} (e.g., with distinct latencies) for any of its transponders.

Observation I captures all instances of operand-dependent hardware resource usage (\S\ref{sec:tool:security-arg}), including prior notions such as \textit{implicit branches}~\cite{jiyong:stt},
e.g., conditional cache accesses for loads (\upath{} variability) that arise due to store and load address-dependent store-to-load forwarding.
Thus, \textbf{our second contribution} is augmenting the aforementioned telecommunications analogy with the notion of a transponder.
\paragraph*{Application: Synthesizing Leakage Contracts}
Observation I also inspires \textbf{our third contribution}: an open-source~\cite{https://github.com/yaohsiaopid/SynthLC} automated approach and tool, called \synthlc{},
which extends \PathFinder{} with a symbolic information flow analysis to support synthesizing 
\textit{a variety of} formally verified microarchitectural \textit{leakage contracts}
from SystemVerilog processor designs. 
Leakage contracts are foundational to hardware side-channel defenses, 
implemented in software~\cite{
vassena:blade, serberus:mosier:2024, zhang2023ultimate, FactLanguage, synthct, narayan2021swivel} or  
hardware~\cite{jiyong:stt, choudhary:2021:spt, Bourgeat:m16, Schwarz:ConTExT, yu:oisa, jiyong:sdo, barber:specshield, dolma}.  
\synthlc{} is the \textit{first automated methodology} for formally verifying hardware adherence to them.

Specifically, \synthlc{} synthesizes a complete set of formally verified microarchitectural \textit{leakage signatures} 
from processor RTL.
Leakage signatures are a novel formalism that we introduce to
capture all relevant features of state-of-the-art leakage contracts, which are not already captured by \upaths{}.
They are effectively function signatures, which
characterize how one or more
transmitters create operand-dependent \upath{} variability for a transponder with respect to (i.e., following) some step (PL) of the transponder's execution.
Fig.~\ref{fig:zero-skip-uhb} depicts a leakage signature, which defines how a \inst{MUL} transmitter creates \upath{} variability for itself (it is also a transponder) as a function of its operands, following its \inst{mulU} execution step.

From \upaths{} and 
leakage signatures, a variety of formally verified leakage contracts can be easily derived.
The canonical \textit{constant-time (\ct{}) contract}, which enumerates a microarchitecture's transmitters and their unsafe operands~\cite{arm:dit,intel:doit,riscv_zkt}, is captured by the inputs to the leakage signature in Fig.~\ref{fig:zero-skip-uhb}.
Overall, \synthlc{} supports synthesizing \textit{six different leakage contracts} from SystemVerilog processor designs (Table~\ref{table:defenses_map}): 
\ct{} contracts plus five bespoke leakage contracts from the literature. Collectively, these contracts
support 
two software defenses---the \textit{\ct{} programming defense}~\cite{FactLanguage, synthct} and the \textit{speculative constant-time (SCT) programming
defense}~\cite{vassena:blade, serberus:mosier:2024, zhang2023ultimate, narayan2021swivel}---and
eight hardware defenses~\cite{Schwarz:ConTExT, barber:specshield, choudhary:2021:spt, Bourgeat:m16, yu:oisa, jiyong:sdo, dolma, jiyong:stt}
against hardware side-channel attacks.

\paragraph*{Case Study: Deploying \synthlc{} on a Processor Core and Cache}
We deploy \synthlc{} on the RISC-V CVA6 core~\cite{cva6}, surfacing
94 unique leakage signatures, 72 transponders, and 26 transmitters. Compared to prior work that analyzes the same design~\cite{deutschmann:upec-do,deutschmann2023scalable}, \synthlc{} finds a novel channel that leaks store and load address operands.

We separately deploy \synthlc{} on the CVA6 L1 data cache and cache controller, making it the first leakage contract verification procedure to analyze a realistic processor cache. 
Beyond uncovering various hardware side channels, our cache experiment showcases
the performance and scalability benefits of converting \synthlc{} into a modular procedure.

%% file: 02-background.tex
\section{Background}
\label{sec:background}

This section gives an informal overview of hardware side-channel attacks (\S\ref{sec:background:hw-sc-attacks}) and defenses (\S\ref{sec:background:hw-sc-defenses}), including descriptions of
six leakage contracts whose implementation in hardware can be formally verified with \synthlc{}.

\subsection{Hardware Side-Channel Attacks}
\label{sec:background:hw-sc-attacks}

In this paper, we study hardware side-channel attacks where a \textit{transmit instruction} (or \textit{transmitter}, in the victim program) modulates a \textit{channel} (hardware resource) in an operand-dependent manner, and a \textit{receiver} (attacker) observes the channel modulation to infer the operand value~\cite{dawg:kiriansky}. We highlight standard assumptions for channels and receivers below.

\paragraph*{Characterizing Channels}
\label{sec:background:telecom:channels}
Many hardware resources have been implicated as channels, e.g., caches~\cite{Osvik:prime+probe+l1d,Yarom:flush+reload+llc13,cache_bleed}, branch predictors~\cite{branchpred_sc, Evtyushkin:BranchScope}, functional units~\cite{andrysco:subnormal,Mult_leaky}, memory ports~\cite{portsmash}, and more~\cite{yan:directories,memjam,xu:controlledchannelattacks,leaky_cauldron,tlb_bleed,drama,pandora:isca:21,augury:2022:vicarte:oakland22}.
Owing to the circumstantial nature of leakage due to particular channels~\cite{intel:side-channel:taxonomy, deutsch2023metior}, prior work informally classifies them
as \textit{stateless} or \textit{stateful}~\cite{intel:side-channel:taxonomy}.
Stateless channel modulations are observable only in very narrow, specific time windows, usually requiring the receiver to be running at the same time as the victim, e.g., a victim may create memory port contention for a receiver during a victim memory access.
Stateful channels may be observed long after they are modulated, e.g., a victim cache access may create a cache miss for a receiver far in the future. 
We refer to stateless/stateful channels as \textit{dynamic}/\textit{static}
to denote formal definitions that we introduce (\S\ref{sec:path-selector}). 

\paragraph*{Receiver Assumptions}
\label{sec:background:telecom:receiver}
Hardware side-channel attacks and defenses are studied under specific \textit{receiver assumptions}, consisting of an \textit{observer model} and \textit{attacker strategy}.

An observer model defines how channel modulations manifest as observations for a specific receiver.
Informally, 
they may be perceived via their (measurable) effects on certain
non-deterministic aspects of program execution, e.g., execution time~\cite{bernstein_aes_attack,exectimeexploit,Osvik:prime+probe+l1d},
resource contention~\cite{branchpred_sc, andrysco:subnormal, portsmash, memjam, xu:controlledchannelattacks, drama, tlb_bleed}, and so on~\cite{perf_counter_sc,powerexploit, acousticexploit, radiationexploit,KocherDPA,sok_em_side_channels}.

An attacker strategy specifies whether the receiver may \textit{passively} or \textit{actively} monitor victim channel modulations~\cite{pandora:isca:21, deutsch2023metior, bourgeat2020casa, spreitzer2017systematic}.
In a passive attack, the receiver monitors victim channel modulations without explicitly interfering with the victim's execution, e.g., by measuring victim execution time.
In an active attack, the receiver explicitly interacts with the channel modulated by the victim, e.g., by
priming and probing shared cache state.
The notion of a transponder enables defining both strategies formally (\S\ref{sec:path-selector}). 

\subsection{Hardware Side-Channel Defenses \& Leakage Contracts}
\label{sec:background:hw-sc-defenses}
The goal of a hardware side-channel defense is to ensure that a specific victim program running on a particular microarchitecture will not leak its private data to some receiver via hardware side channels.
Towards this goal, state-of-the-art defenses assume the availability of microarchitectural \textit{leakage contracts},
which characterize implementations' transmitters.

\paragraph*{The Canonical Leakage Contract}
The most widely-adopted leakage contract, the \textit{constant-time (\ct{}) contract}, enumerates a microarchitecture's transmitters and their unsafe (``leaky'') operands. 
Nascent ISA leakage contracts fall into this category~\cite{arm:dit, intel:doit, riscv_zkt}. 
Given a \ct{} contract, a hardware side-channel defense can ensure that secrets \textit{never} reach the unsafe operands of transmitters when programs execute. This strategy is embodied in the
\textit{CT programming defense}~\cite{FactLanguage, synthct}, the 
gold standard software defense for protecting the \textit{architectural} executions of cryptopgraphic code~\cite{openssl, libsodium, hacl, mbedtls} 
from hardware side-channel attacks.
%
Some 
software~\cite{vassena:blade, serberus:mosier:2024, zhang2023ultimate, narayan2021swivel} and 
hardware~\cite{barber:specshield, Schwarz:ConTExT} defenses against
\textit{transient execution attacks}---which exploit hardware faults and mis-predictions to \textit{steer} secrets towards the unsafe operands of \textit{transient} (bound-to-squash) transmitters~\cite{spectre,canella2019systematic}---extend this strategy to target programs'
\textit{speculative} executions as well.
We refer to software variants of such defenses as \textit{speculative constant-time (SCT) programming defenses}~\cite{vassena:blade}.

\paragraph*{Five Bespoke Leakage Contracts}
Today, the most performant
defenses against transient execution attacks are implemented in hardware~\cite{jiyong:stt,choudhary:2021:spt,Bourgeat:m16,  yu:oisa,jiyong:sdo,dolma}.
Instead of \ct{} contracts~\cite{Bourgeat:m16, jiyong:stt, jiyong:sdo} (or in addition to them~\cite{yu:oisa, dolma, choudhary:2021:spt}), these defenses adopt bespoke leakage contracts that are much finer grained.
We consider five such fine-grained leakage contracts in this paper.
Table~\ref{table:defenses_map} in \S\ref{sec:synth-lc-approach} shows how features of these five contracts, described below for each of the six hardware defenses they enable, and the \ct{} contract map onto the various components of our proposed leakage signatures.

MI6~\cite{Bourgeat:m16} defends enclaves from hardware side-channel attacks via two mechanisms.
First, it requires identifying \textit{dynamic (stateless---\S\ref{sec:background:hw-sc-attacks}) channels} that arise due to transmitter operand-dependent hardware \textit{resource contention}.
Data-independent scheduling logic is implemented for impacted resources.
Second, it requires identifying \textit{static (stateful) channels} in order to implement a purge instruction, which flushes  relevant microarchitectural states, and partitioning schemes.

OISA~\cite{yu:oisa} 
attaches secrecy labels to architectural state and detects if a transmitter's unsafe operand, as defined in a \ct{} contract, is ever passed secret data at runtime.
It enables
transmitters that exhibit execution variability
due to \textit{input-dependent arithmetic units} (like \inst{MUL} in \S\ref{sec:intro:thispaper}) to safely process secrets as follows.
First, the designer identifies arithmetic units that may be occupied by a transmitter for an operand-dependent number of cycles.
Second, control logic is added to enforce operand-independent execution for the unit whenever a transmitter arrives with a secret-labeled unsafe operand.


STT~\cite{jiyong:stt} requires identifying all channels and classifying them as 
\textit{explicit channels} 
and \textit{implicit channels}.
Explicit and implicit channels, respectively, arise when transmitters' operand-dependent hardware resource usage affects their own execution behavior and the execution behavior of other instructions.
Transmitters that modulate explicit channels are not permitted to execute with (potentially) secret operands.
To block implicit channels, designers must also identify
\textit{explicit branches} or \textit{implicit branches}. Explicit branches are architectural control-flow instructions, like conditional branches. Implicit branches are instructions whose execution behavior depends on the operands of other transmitters.
Finally, implicit channels are categorized as \textit{prediction-based} or \textit{resolution-based}.
To block prediction-based and resolution-based channels, respectively, the designer must identify predictor structures that are updated by explicit or implicit branches and the points at which these branches resolve their predictions.

SDO~\cite{jiyong:sdo} extends STT by optimizing its explicit channel defense. 
First, the designer identifies transmitters that modulate explicit channels. 
To enable safe speculative execution of these transmitters with (potentially) secret operands, the designer creates a number of
data-independent execution path variants for each. These so-called \textit{data-oblivious variants} are derived from the set of realizable microarchitectural execution paths for each transmitter on the baseline design.
Then, the designer
adds control logic to select which path a transmitter should take based on public predictor state.

Dolma~\cite{dolma} requires a \ct{} contract to delay the execution of transmitters until they become non-speculative. To improve performance, Dolma has several additional requirements. First, the designer must identify \textit{variable-time micro-ops} and any \textit{contention-based dynamic channels} they create. 
Second, \textit{inducive micro-ops} that exhibit execution variation as a function of \textit{resolvent micro-ops'} (transmitters') 
operands must be flagged, along with the \textit{prediction resolution point} during the inducive micro-op's execution at which this variation arises.
Lastly, \textit{persistent state modifying micro-ops} (transmitters that modulate static channels) must be identified and their persistent state updates delayed until they become non-speculative.

SPT~\cite{choudhary:2021:spt} and STT have the same fine-grained leakage contract and 
differ only in their policy for when it safe to declassify (mark public) data.
STT declassifies data once it is not a function of speculatively accessed data.
SPT declassifies data once it is inevitable that it will be, or has been, transmitted architecturally; so, SPT additionally requires a \ct{} contract.

%% file: 03-upath-formalism.tex
\section{\texorpdfstring{\PathFinder{}}{SynthLC} Approach: Formalizing Microarchitectural Execution Paths}
\label{sec:upath_formal}

Our first contribution is an automated approach and tool, called
\PathFinder{},\footnote{``\PathFinder{}'' indicates that it finds \textit{multiple} (\textsc{m})  \upaths{} per instruction.} 
for uncovering a complete set of formally verified \textit{microarchitectural execution paths} (\upaths{}) for each instruction  implemented on a SystemVerilog processor design.
In this section, we explain the two key technical advances
that enable such an analysis: an extension to the \uhb{} graph formalism from prior work (\S\ref{sec:uhb-timing}) and a novel mapping of \uhb{} nodes onto RTL signals (\S\ref{sec:pls}). We establish the need for both with a motivating example
(\S\ref{sec:op-packing-example}). A discussion of \PathFinder{}'s implementation details is deferred until \S\ref{sec:both-tool}.


\subsection{Motivating Example: Operand Packing}
\label{sec:op-packing-example}


Consider an extension to the RISC-V CVA6 CPU~\cite{cva6}
that we verify in \S\ref{sec:case-study}, called CVA6-OP. The baseline CVA6 design may fetch up to two compressed instructions or one uncompressed instruction per cycle, 
but only one instruction can be sent to decode per cycle. CVA6-OP is identical to 
CVA6, except that
the ALU has been modified to support \textit{operand packing}~\cite{brooks:operand-packing}, 
and up to \textit{two} instructions can be sent to decode per cycle.
In particular, if a pair of (concurrently) decoded instructions perform identical ALU operations \textit{and} feature narrow width operands (32 bits or less, i.e., the upper halves of their 64-bit operands contain all zero bits), they will be packed into a single operation and executed together.


Suppose we want to formally evaluate the vulnerability of CVA6-OP to hardware side-channel attacks with the \textit{Check} tools (i.e., with \textit{CheckMate}~\cite{checkmate}).
As discussed in \S\ref{sec:intro}, we first need to acquire an abstract axiomatic model of the microarchitecture, called a \uspec{} model.

The enabler for \textit{Check}-based hardware security verification is the fact that \uspec{} models define how instructions may exhibit variable hardware resource usage when they run on a specific hardware implementation (e.g., a load may experience a cache hit or miss~\cite{checkmate}). That is, they specify the various \textit{microarchitectural execution paths} (\upaths{}---our term) that may be exhibited by each instruction type (e.g., a \inst{LD}, \inst{MUL}, \inst{ADD}, etc.) on the design.
A \uspec{} model encodes distinct \upaths{} as distinct \textit{microarchitectural happens-before} (\uhb{}) graphs (Fig.~\ref{fig:add-examples}). A node in a \uhb{} graph represents a microarchitectural event, such as a dynamic program
instruction (column label) updating a particular set of hardware state elements (row label) during its execution~\cite{rtl2uspec:hsiao:2021}. Directed edges denote happens-before relationships~\cite{lamport:happensbefore}.

\input{fig-add-path}

A \uspec{} model for CVA6-OP \textit{should} capture the fact that \inst{ADD}s may exhibit two distinct \upaths{}, depending on whether they are packed or not. 
However, prior work~\cite{pipecheck, ccicheck, coatcheck, tricheck, rtlcheck, pipeproof, checkmate} would represent both of these scenarios with the same \upath{}---the one in Fig.~\ref{fig:add_standard}. This is because the packed versus unpacked \upaths{} differ according to how long an \inst{ADD} spends in the decode stage (one or two cycles), i.e., how many times the \inst{ID} node appears. The \uhb{} formalism has not been used to express repeated instances of the same \uhb{} node.

Even if we resolve this minor modeling issue, the only approach and tool for synthesizing formally verified \uspec{} models from processor RTL, called \rtltouspec{}, cannot uncover more than one \upath{} per implemented instruction (\S\ref{sec:intro}).
Beyond missing repeated instances of the same node, \rtltouspec{}'s mapping of \uhb{} nodes to RTL signals precludes identifying nodes that appear in some \upaths{} but not others. We address these limitations in \S\ref{sec:uhb-timing} and \S\ref{sec:pls}, respectively.

\subsection{Extending \uhb{} Graphs with Cycle-Accurate Timing}
\label{sec:uhb-timing}
We extend the \uhb{} graph formalism from prior work~\cite{pipecheck, ccicheck, coatcheck, tricheck, rtlcheck, pipeproof, checkmate, rtl2uspec:hsiao:2021} with \textit{cycle-accurate} timing information. 
In particular, \textit{for all \upaths{} in this paper} (except in Fig.~\ref{fig:add_standard}), a \uhb{} node represents an instruction updating a particular set of state elements \textit{in a specific cycle}, while an edge represents a \textit{one-cycle} happens-before relationship.
This extension enables \uhb{} graphs to express
that an instruction may update the same set of state elements
in multiple (consecutive or non-consecutive) cycles, which is needed to represent real designs, e.g., ones where an execution unit has variable latency (like CVA6-MUL in Fig.~\ref{fig:zero-skip-uhb}) or throughput (like CVA6-OP in Fig.~\ref{fig:add-examples}).


Using our cycle-accurate \uhb{} graph notation, the
\upaths{} in Figs.~\ref{fig:add_path_1} and \ref{fig:add_path_2} depict distinct (concrete) executions of an \inst{ADD} on CVA6-OP
that distinguish
when the \inst{ADD} is packed or not, respectively.
We use row label \inst{Row(n)} to denote the \inst{n}-th update to the set of state elements referred to by \inst{Row}.

We use a special notation to summarize $l$
\textit{consecutive} updates to the same set of state elements.
Specifically, a pair of row labels \inst{Row(1)} and \inst{Row(}$l$\inst{)} denote the first and last ($l$-th) consecutive updates to the state elements referred to by \inst{Row}. Node labels specify the value of $l$ (which may be execution dependent) for a particular concrete \upath{}.
For example, in Fig.~\ref{fig:add_path_2}, the node at \inst{ID(}$l$\inst{)} denotes an \inst{ADD}'s second consecutive update to the state elements referred to by \inst{ID}.
Dashed edges that relate a \inst{Row(}$1$\inst{)} node to its corresponding \inst{Row(}$l$\inst{)} node represent a totally ordered (by \uhb{} edges) sequence of 
$l-2$ 
nodes, ordered after \inst{Row(1)} and before \inst{Row(}$l$\inst{)}, with no other outgoing edges. When $l=2$ (as in Fig.~\ref{fig:add_path_2} for \inst{ID(}$l$\inst{)}), the dashed edge represents a normal (solid) \uhb{} edge.

In every \upath{} in this paper---each of which was synthesized from CVA6~\cite{cva6} using \PathFinder{} (or, in the case of Figs.~\ref{fig:zero-skip-uhb}~and~\ref{fig:add-examples}, adapted from synthesized \upaths{})---\inst{IF} represents the first set of state elements that an instruction updates upon being fetched, and
\inst{scbCmt} represents those updated upon commit.
Thus,
one can deduce that a set of state elements \inst{S} (row label) is updated in cycle $t$ of an instruction's execution, along some \upath{}, 
if $t$ is longest sequence of \uhb{} edges 
from the node at $\mathtt{IF}$ to the node at \inst{S}, accounting for implicit nodes and edges due to consecutive state updates.
An instruction's overall latency is given by the longest sequence of \uhb{} edges from the node at $\mathtt{IF}$ to the node at $\mathtt{scbCmt}$, e.g., four (Fig.~\ref{fig:add_path_1}) or five (Fig.~\ref{fig:add_path_2}) cycles for a packed or non-packed \inst{ADD} on CVA6-OP, respectively.

In summary, our cycle-accurate \upath{} abstraction provides a precise accounting of an instruction's hardware resource  usage {(i.e., state updates)} in time and space. 

\subsection{Recognizing \uhb{} Nodes with Performing Locations}
\label{sec:pls}
\PathFinder{} uses
static netlist analysis, 
linear temporal logic (LTL)~\cite{pnueli:ltl,manna:ltl-textbook} property generation (from property templates), and model checking~\cite{baier:principles_of_model_checking, clarke:model_checking}
to uncover all \upaths{} for each implemented instruction on a given SystemVerilog processor design (\S\ref{sec:both-tool}). Importantly, it requires that a model checker be able to recognize the various components of a \upath{}---\uhb{} nodes and edges---when exploring instructions' executions. As one-cycle happens-before relationships, edges are readily expressible in LTL syntax. However, nodes must be explicitly conceptualized in terms of RTL signals.


Recall that a \uhb{} node represents an instruction updating a particular set of design states in a given cycle (\S\ref{sec:uhb-timing}). Thus, a model checker recognizing a \uhb{} node equates to it detecting what state elements are updated in some execution cycle and attributing these updates to a particular in-flight instruction. At first glance, this task appears highly challenging, given that modern processors execute numerous instructions concurrently and out-of-order. However, we observe that the same mechanisms that enable a processor to orchestrate
instructions' write access to design states can also be leveraged by a model checker to recognize \uhb{} nodes.

In particular, we find that a subset of \textit{finite state machines} (FSMs) within a processor's control path, which we call \textit{micro-op FSMs} (\ufsms{}), govern instructions' state updates throughout their execution, from the time they are fetched until possibly after they commit (e.g., when stores update cache state).
A \ufsm{} is a tuple $\langle \mathtt{iir}, \mathtt{vars} \rangle$, where \inst{iir} is an \textit{instruction identifying register} (IIR), which holds a unique \textit{instruction identifier} (IID) for an in-flight instruction, and \inst{vars} is a collection of state elements, which encode the \ufsm{}'s \textit{state variables}. Example IIDs include program counters (PCs)~\cite{reid:isaformal,rtl2uspec:hsiao:2021}, reorder buffer (ROB) or scoreboard (SCB) identifiers, and memory transaction identifiers.
An in-flight instruction acquires a \ufsm{} by placing one of its IIDs in the \ufsm{}'s \inst{iir}, progresses through various \ufsm{} states (i.e., concrete valuations of its \inst{vars}), and then releases the \ufsm{} (i.e., by setting its \inst{vars} to an 
\textit{idle} state).
The valuation of a \ufsm{}'s \inst{vars}
in a given cycle grants
the instruction whose IID is contained in its \inst{iir} exclusive write access to a particular subset of design states.

\input{fig-svfile}
Leveraging \ufsms{}, we conceptualize \uhb{} nodes in terms of RTL signals as follows.
First, we define the notion of a \textit{performing location} (PL)---similar to a pipeline stage, but more granular---as a $\langle \mu\mathtt{fsm}, \mathtt{state} \rangle$ tuple, 
where $\mu\mathtt{fsm}$ is a \ufsm{} (i.e., an $ \langle \mathtt{iir}, \mathtt{vars} \rangle$ tuple), and \inst{state} denotes a valid, non-idle valuation of $\mu\mathtt{fsm}$'s \inst{vars}. 
Hence, the set of all PLs for a processor implementation denotes the set of all valid, non-idle states across all of the design's \ufsms{}.
Next, we say that ``an instruction $i$ \textit{visits} (i.e., \textit{occupies}) PL $\langle \mu\mathtt{fsm}, \mathtt{state} \rangle$ some cycle,'' if at the start of that cycle, $\mu\mathtt{fsm}$'s \inst{iir} contains an IID of $i$ and $\mu\mathtt{fsm}$'s \inst{vars} is set to \inst{state}.
Like a pipeline stage, a PL can be occupied by a single instruction at a time, whose state updates (which take effect at the start of the next cycle) it encapsulates.
Unlike a pipeline stage, an instruction may visit multiple PLs in the same cycle.
Finally, we direct a model checker to recognize a \uhb{} node during an instruction's execution on a microarchitecture when it detects the instruction visiting a particular PL in a given cycle.
Hence, in all \upath{} figures in this paper, row labels (ignoring parentheticals) denote PLs.

We require the designer to identify all signals that comprise \ufsms{} (i.e., their IIR and state variable components) in the input design (\S\ref{sec:tool:metadata}). 
\PathFinder{} uses these signals to uncover all PLs for the design and then all ways in which they can be assembled into valid \upaths{} for instructions.
Specifying \ufsms{} turns out to be straightforward, since they are already in use to support functional verification efforts~\cite{ho:global-fsms, ho:global-fsms-thesis}.
Fig.~\ref{fig:iir_pcr} illustrates the close proximity of \ufsms{}' IIR and state variable
signal definitions in CVA6~\cite{cva6}.

%% file: fig-add-path.tex
\begin{figure}[t]
  \centering
   \begin{subfigure}{0.32\linewidth}
       \centering
     \includegraphics[height=3.5cm]{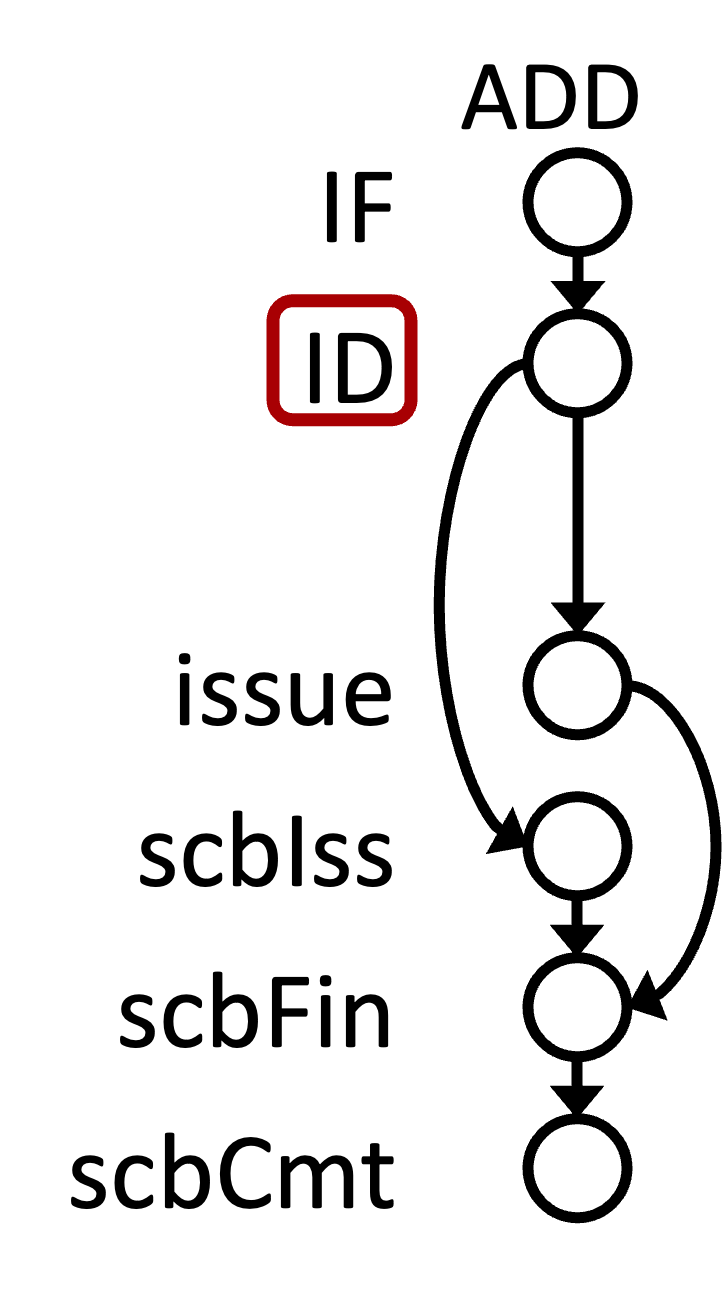}
       \caption{\textit{(Non-)packed} \upath{}}
       \label{fig:add_standard}
   \end{subfigure}
   \hfill
    \begin{subfigure}{0.32\linewidth}
      \centering
      \includegraphics[height=3.5cm]{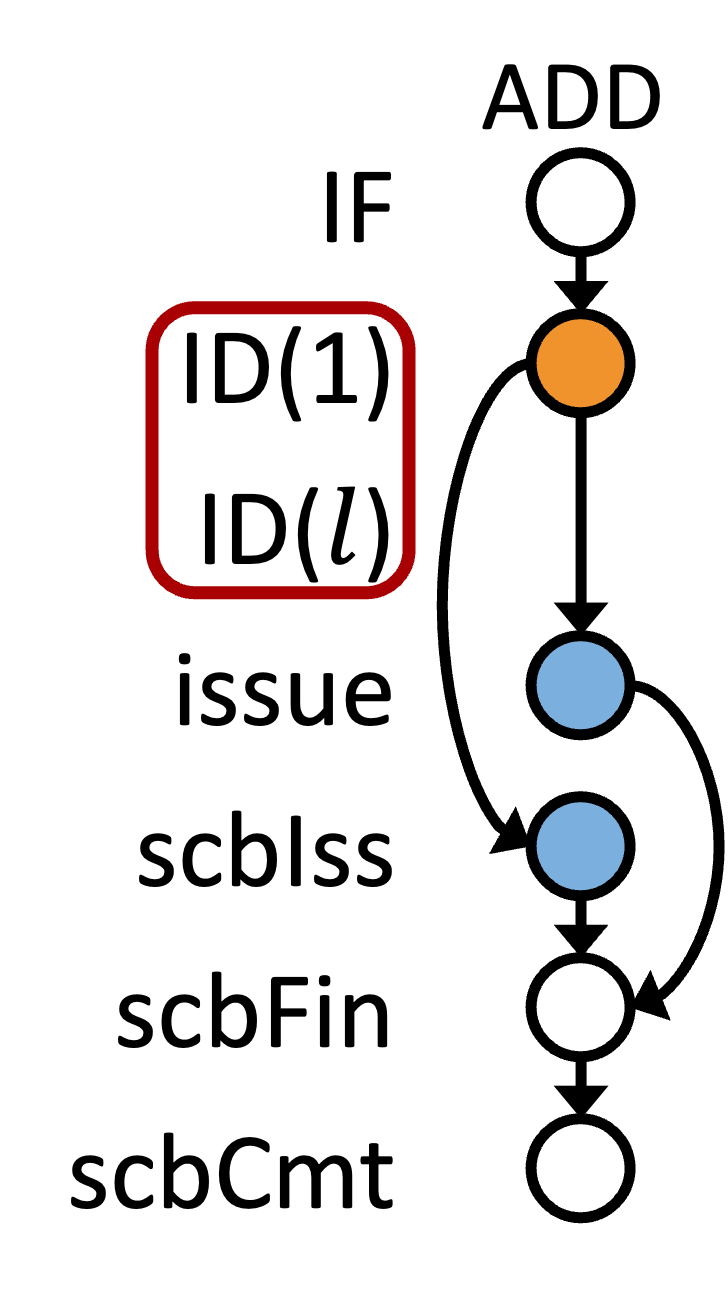}
      \caption{\textit{Packed} \upath{}}
      \label{fig:add_path_1}
  \end{subfigure}
  \hfill
  %
  %
  \begin{subfigure}{0.32\linewidth}
      \centering
    \includegraphics[height=3.5cm]{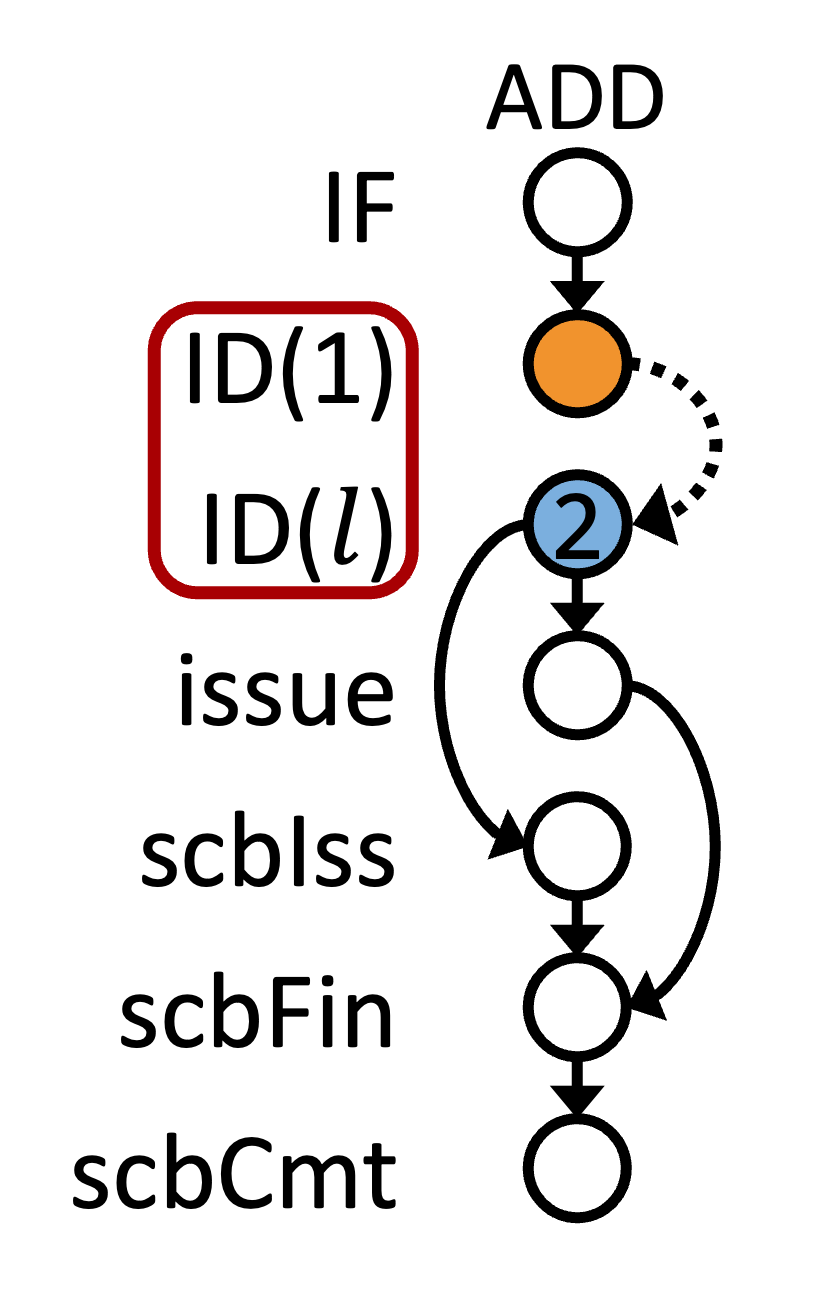}
      \caption{\textit{Non-packed} \upath{}}
      \label{fig:add_path_2}
  \end{subfigure}
      \caption{\inst{ADD} \upath{}s on CVA6-OP (\S\ref{sec:upath_formal}), using standard \uhb{} graphs from prior work (\subref{fig:add_standard}) and our new cycle-accurate \uhb{} graphs (\subref{fig:add_path_1}, \subref{fig:add_path_2}).
      Row({\inst{1}/\llt{}}): 1st/\llt{}-th visit to Row.
      Node label: value of \llt{} for \upath{}. 
      }
      \label{fig:add-examples}
\end{figure}

%% file: fig-svfile.tex
\begin{figure}[!t]
\begin{minipage}{\linewidth}
\begin{lstlisting}[style=svaListing]
// store_unit.sv
  enum logic {...} state_d, state_q;              // vars0 
  logic [TRANS_ID_BITS-1:0] trans_id_n, trans_id_q;// iir0
^\svaplus{+}^ logic [riscv::VLEN-1:0] st_pc_n, st_pc_q;        // pcr0 
// load_store_unit.sv
  assign lsu_req_i = { 
  lsu_valid_i, fu_data_i.trans_id, ...   // (vars1, iir1)
^\svaplus{+}^ , pc_i };                              // pcr1
\end{lstlisting}
\end{minipage}
\vspace{-5pt}
  \caption{Close proximity of \ufsms{}' \inst{iir} and \inst{vars} components (\S\ref{sec:pls}). PCRs are added (+) near IIRs that do not hold PCs (\S\ref{sec:both-tool}).}
\label{fig:iir_pcr}
\end{figure}

%% file: 03-tspec.tex
\section{\texorpdfstring{\synthlc{}}{SynthLC} Approach: Applying Multi-\texorpdfstring{\upath{}}{upath} Synthesis to Hardware Security Verification}
\label{sec:synth-lc-approach}
In designing \PathFinder{}, we make an important observation: an instruction that exhibits \upath{} variability (${>}1$ \upath{}) on some processor implementation strongly indicates the presence of a microarchitectural side channel in the design. Based on this observation, we develop \synthlc{}, the first automated approach and tool for formally verifying hardware adherence to microarchitectural leakage contracts (\S\ref{sec:background:hw-sc-defenses}).

This section presents the \synthlc{} approach; a presentation of its implementation as a tool is deferred until \S\ref{sec:both-tool}. 
We first formalize instances of instruction \upath{} variability using the notion of a  \textit{decision} (\S\ref{sec:decisions}). Second, we show how an instruction's decisions can be attributed to the outputs of \textit{path selector functions} in the microarchitecture (\S\ref{sec:path-selector}). If a path selector function output depends on some instruction's operands, it is a \textit{leakage function}, and said instruction is a transmitter. Instructions whose decisions can be attributed to leakage functions are \textit{transponders}. From leakage functions, we derive \textit{leakage signatures}---a unifying formalism that captures all relevant features of six leakage contracts from prior work (summarized in \S\ref{sec:background:hw-sc-defenses} and Table~\ref{table:defenses_map}), which are not already captured by \upaths{} (\S\ref{sec:leakage-signatures}). 
\synthlc{} conducts leakage contract verification by synthesizing a comprehensive set of
leakage signatures from processor RTL
(\S\ref{sec:both-tool}).

We present the concepts above using a more complex illustrative example involving store-to-load stalling (\S\ref{sec:store-to-load-stall}).



\input{fig-perinst-paths}

\subsection{Illustrative Example: Store-to-Load Stalling}
\label{sec:store-to-load-stall}
Consider two \upaths{} for a load (\inst{LD}) executing on the RISC-V CVA6 CPU~\cite{cva6}, shown in Fig.~\ref{fig:ld-upaths}. Following its visit to the \inst{issue} PL (orange node), a \inst{LD} exhibits one of two \textit{decisions}. It stalls (right \upath{}), progressing to \inst{LSQ} and \inst{ldStall} (blue nodes), if the page offset of its address operand matches that of any older pending store (\inst{ST}) in the speculative or committed store buffers (STBs)---its \textit{path selector function}. Else, it completes (left), progressing to \inst{ldFin} (blue node).

Hardware side-channel defenses (\S\ref{sec:background:hw-sc-defenses}) must ensure that stores (\textit{transmitters}) do not leak their (private) address operands to a receiver who observes the \upaths{} of executing loads (\textit{transponders}), e.g., by timing their execution latency---five or nine cycles for the left or right \upath{}, respectively, in Fig.~\ref{fig:ld-upaths}.
CT contracts would declare store address operands as unsafe for processing secrets.
STT, SDO, and SPT would flag such a load as an implicit branch, while Dolma would categorize the load/store as inducive/resolvent micro-ops.
OISA would specify that store address operands are unsafe or require hardware mechanisms avoid this channel for stores with secret operands.
MI6 declares this sort of leakage as out-of-scope.

\subsection{Formalizing Instances of \texorpdfstring{\upath{}}{upath} Variability with Decisions}
\label{sec:decisions}
We propose the notion of a \textit{decision} to characterize specific variations across the different \upaths{} of a particular instruction on a  microarchitecture (e.g., a \inst{LD} on CVA6).

Suppose $\boldsymbol{\mu \textsc{\textbf{path}}}^{I}_M$ is the set of all possible \upaths{} that a dynamic instance of instruction $I$ (e.g., $I = \mathtt{LD}$) can exhibit when it runs on microarchitecture \M{}. 
Informally, ``a decision made by $I$ on \M{}'' is a tuple $(src, \mathbf{dst})$---where $src$ is a single \textit{decision source PL} (or \textit{decision source})
and $\mathbf{dst}$ is a set of \textit{decision destination PLs} (or \textit{decision destinations})---that
pinpoints a divergence between a pair of
$I$'s \upaths{} on \M{}.
Orange/blue nodes in \uhb{} graph figures throughout the paper denote some, \textit{but not all} (for clarity of presentation),
decision sources/destinations.
Suppose $\mathbf{d}^{I}_M$ is the set of all decisions that $I$ can make on \M{}.
Then, $(src, \mathbf{dst}) \in \mathbf{d}^{I}_M$ if and only if: For some $p, p' \in \boldsymbol{\mu \textsc{\textbf{path}}}^{I}_M$, $I$ visits $src$ in $p$ one cycle before it visits exactly the PLs in $\mathbf{dst}$, and  $I$ visits $src$ in $p'$ one cycle before it visits a \textit{different} set of PLs than exactly those in $\mathbf{dst}$.
By $\mathbf{src}^I_M$ we denote the set of all decision sources across all of $I$'s decisions on \M{}.

For example, given the \upaths{} in Figs.~\ref{fig:add_path_1}~and~\ref{fig:add_path_2}, and considering the orange and blue colored nodes exclusively (for brevity), \inst{ADD} has decision sources and decisions:
$$\mathbf{src}^\mathtt{ADD}_\mathrm{CVA6\hbox{-}OP} = \Bigl\{\mathtt{\color{myorange} ID}\Bigl\}$$ 
\vspace{-10pt}
$$\mathbf{d}^\mathtt{ADD}_\mathrm{CVA6\hbox{-}OP} = \Bigl\{(\mathtt{\color{myorange} ID}, \{\mathtt{\color{myblue} issue}, \mathtt{\color{myblue} scbIss}\}), (\mathtt{\color{myorange} ID}, \{\mathtt{\color{myblue} ID}\})\Bigl\}.$$
Similarly, given Fig.~\ref{fig:ld-upaths} and considering only colored nodes, \inst{LD} has decision sources and decisions: $$\mathbf{src}^\mathtt{LD}_\mathrm{CVA6} = \Bigl\{\mathtt{\color{myorange} issue}\Bigl\}$$ 
\vspace{-10pt}
$$\mathbf{d}^\mathtt{LD}_\mathrm{CVA6} = \Bigl\{(\mathtt{\color{myorange} issue}, \{\mathtt{\color{myblue} ldFin}\}), (\mathtt{\color{myorange} issue}, \{\mathtt{\color{myblue} LSQ}, \mathtt{\color{myblue} ldStall}\})\Bigl\}.$$
Note that in practice, decisions are defined with respect to PLs, irrespective of how many times they have been visited. For example, in Fig.~\ref{fig:zero-skip-uhb}, \inst{\color{myorange} scbIss} is a decision source for \inst{MUL} on CVA6-MUL; it may be followed by decision destinations 
$\{$\inst{\color{myblue} scbIss, mulU}$\}$, 
$\{$\inst{\color{myblue} scbIss}$\}$,
or $\{$\inst{\color{myblue} scbCmt}$\}$.

\subsection{Selecting a Decision with a Path Selector Function}
\label{sec:path-selector}

Suppose $i_I$ is a dynamic instance of instruction $I$ that visits decision source $src \in \mathbf{src}^I_M$ during its execution on microarchitecture \M{}. Which decision $i_I$ exhibits with respect to $src$---i.e., which decision destination(s) $i_I$ visits one cycle after visiting $src$---is determined by a \textit{path selector function} in hardware. In particular, during the cycle in which $i_I$ visits $src$, a path selector function is queried to determine where $i_I$ will progress to next. We use \inst{I\_src} to denote a path selector function that is queried when an instruction of type $I$ visits decision source $src$; it returns a set of decision destinations.

\input{fig-f-leak}
Fig.~\ref{fig:leakage_func_examples}
shows example path selector functions for CVA6-OP and CVA6~\cite{cva6}.
Path selector functions may have \textit{explicit inputs} that are provided in the function argument list and \textit{{implicit inputs}} 
that are not.
Explicit inputs are instructions whose operands appear in the function body, capturing how architectural state influences \upath{} variability.
Implicit inputs are any other microarchitectural structures whose contents are accessed in the function body, capturing how microarchitectural state influences \upath{} variability.

Moreover, each explicit input to a leakage signature has a type that captures both its \textit{instruction type} (opcode/function) and particular \textit{runtime conditions} (encoded with \inst{N}, \inst{D}, or \inst{S} superscripts and \inst{O} or \inst{Y} subscripts---detailed below) that must be satisfied for the leakage signature to be applicable.
In particular, with respect to a dynamic instruction $i_I$ of type (opcode/function) $I$ that visits 
$src \in \mathbf{src}^I_M$ on microarchitecture \M{} 
and queries 
path selector function \inst{I\_src},
each explicit input $i_T$ (an instruction with opcode/function $T$) is typed as:

\begin{itemize}
    \item \textit{Intri\underline{n}sic} ($T^\mathtt{N}$ $i_T$), if $i_I = i_T$, i.e., $i_T$ is the instruction currently visiting $src$.
    \item \textit{\underline{D}ynamic}  ($T^\mathtt{D}_\mathtt{{O{\color{black}|Y}}}$ $i_T$), if $i_T$ {\color{black} is older (\inst{O}) / younger (\inst{Y}) than 
    $i_I$ (in program order),
    and $i_T$} must be in-flight (visiting some PL) when $i_I$ visits $src$ for it to influence \inst{I\_src}'s output.
    Notably, when $i_T$ is younger than $i_I$, $M$ can be susceptible to speculative interference attacks~\cite{Behnia:speculative-interference}.
    \item \textit{\underline{S}tatic} ($T^\mathtt{S}$ $i_T$), otherwise.
\end{itemize}

\paragraph*{Illustrative Example: \inst{LD\_issue} Path Selector Function}
For our store-to-load stalling example (\S\ref{sec:store-to-load-stall}), the path selector function \inst{LD\_issue} in Fig.~\ref{fig:leakage_func_examples} is queried to select a decision for a dynamic \inst{LD} instruction $i_\mathtt{LD}$ during any cycle in which it visits the \inst{issue} PL.
Its explicit inputs \inst{LD^N} $i0$ and \inst{ST^D_{\color{black}O}} $i1$ indicate that $i_\mathtt{LD}$'s decision at decision source \inst{issue} is a function of its own operands and those of another older in-flight \inst{ST}, respectively.
Its outputs are one of two sets of destination PLs: \inst{\{ldStall, LSQ\}} (stall \upath{}) or \inst{\{ldFin\}}.
The function body shows that the output of \inst{LD\_issue} specifically depends on the \textit{address} operands of $i_\mathtt{LD} = i0$ and $i1$.

A receiver that can determine which \upath{} $i_\mathtt{LD}$ exhibits relative to $\mathtt{issue}$, learns \inst{LD\_issues}'s return value.
Since the return value depends on the address operands of explicit inputs $i_\mathtt{LD} = i0$ and $i1$, such a receiver may infer their values.

\paragraph*{Leakage Functions and  Transponders}
At least one operand of an explicit input to a path selector function may leak its value to a receiver that observes the function output.
So, we call a path selector function with at least one explicit input a \textit{leakage function}; explicit inputs are \textit{transmitters}.
A receiver observes a path selector function output as a
\upath{} decision of the instruction that queried it.
We call an instruction whose perceived \upath{} variability leaks transmitter operands a \textit{transponder}, extending the  telecommunications analogy for characterizing hardware side-channel attacks (\S\ref{sec:background:hw-sc-attacks}).

To summarize, a leakage function $\mathtt{dst~P\_src } (i_T, i_{T'}, ...)$ characterizes a microarchitectural side channel, mapping the operand space(s) of transmitters $T, T', ...$ that modulate it onto the observation space of a receiver that observes it.
Assuming a receiver that observes \upaths{} of executing instructions,
the observation space consists of the set of distinct decisions 
that \textit{transponder} \inst{P} may exhibit relative to decision source \inst{src}.

\input{table-hw-defense-2}
\paragraph*{Formally Characterizing Channels}
Notably, leakage functions enable formally defining static versus dynamic channels and passive versus active attacks (\S\ref{sec:background:telecom:receiver}).

We call a channel (leakage function) static iff it is modulated by a static transmitter. We call a channel dynamic iff it is modulated by an intrinsic or dynamic transmitter. A channel may be static \textit{and} dynamic, e.g., consider a leakage function whose output decides whether or not a transponder \inst{LD} stalls on a cache access. The decision may depend on another dynamic \inst{LD^D} that contends for the same read port \textit{or} another static \inst{LD^S} that evicted \inst{LD}'s cache line, causing a cache miss. \synthlc{} discovers this scenario when analyzing the CVA6 cache (\S\ref{sec:case-study}).

By definition, a transmitter is an instruction in a victim program (\S\ref{sec:background:hw-sc-attacks}). 
In a passive attack, the transponder is a victim instruction, which the attacker passively monitors. In an active attack, the transponder is a receiver instruction.





\subsection{Unifying Leakage Contracts with Leakage Signatures}
\label{sec:leakage-signatures}
We propose a unifying formalism, called a \textit{leakage signature}, to capture all relevant features of six state-of-the-art leakage contracts, summarized in \S\ref{sec:background:hw-sc-defenses}, which are not already captured by \upaths{}.
A leakage signature is a leakage function that is restricted to the yellow-highlighted components of Fig.~\ref{fig:leakage_func_examples}: transponder and decision source (function name), typed
transmitters (explicit inputs), unsafe transmitter arguments (in the function body), 
and decision destinations (return values).

Table~\ref{table:defenses_map} shows how various components of the six leakage contracts in \S\ref{sec:background:hw-sc-defenses} can be derived from \upaths{} ($\mu$) and leakage signatures. These contracts do not consider notions of decision destinations nor do they explicitly distinguish older versus younger dynamic transmitters, so 
the table omits these details.

Consider the five leakage contract components shared by STT~\cite{jiyong:stt}, SDO~\cite{jiyong:sdo}, and SPT~\cite{choudhary:2021:spt}.
Each can be derived from the checked (\chkm) leakage signature components in Table~\ref{table:defenses_map} as follows.
\textit{Explicit channels} denote sources of \upath{} variability (\inst{src}) for intrinsic transmitters ($T^\mathtt{N}$), which are transponders (\inst{P}) by definition, as a function of their  arguments (\inst{a}).
\textit{Implicit channels} are sources of \upath{} variability (\inst{src}) for transponders (\inst{P}) as a function of dynamic ($T^\mathtt{D}$) or static ($T^\mathtt{S}$) transmitters'  arguments (\inst{a}).
\textit{Implicit branches} are transponders (\inst{P})
that exhibit \upath{} variability due to dynamic ($T^\mathtt{D}$) or static ($T^\mathtt{S}$) transmitters'  arguments (\inst{a}).
\textit{Prediction-based channels} and \textit{resolution-based channels} manifest as sources of \upath{} variability (\inst{src}) for transponders (\inst{P}) due to dynamic ($T^\mathtt{D}$) and static ($T^\mathtt{S}$) transmitters'  arguments (\inst{a}), respectively.

\label{sec:tspec:pct_usage}
\label{sec:tspec:leakage-functions}

%% file: fig-perinst-paths.tex
\begin{figure}[!t]
    \centering
    \begin{subfigure}[t]{0.25\linewidth}
    \centering
    \includegraphics[height=5.5cm]{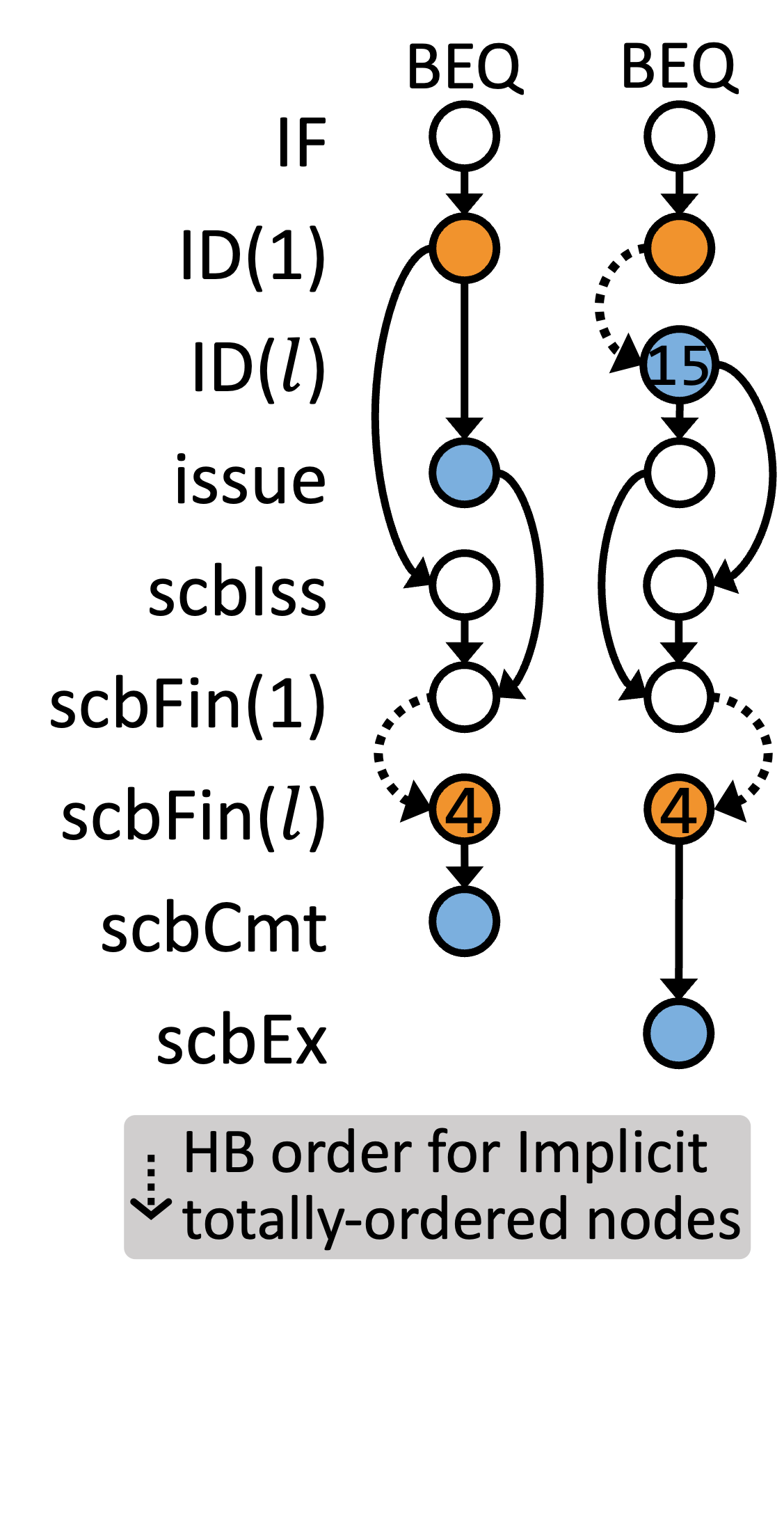}
    \caption{\inst{BEQ} \upaths{}}
     \label{fig:beq-upaths}
    \end{subfigure}
    \hfill
    \begin{subfigure}[t]{0.32\linewidth}
    \centering
    \includegraphics[height=5.6cm]{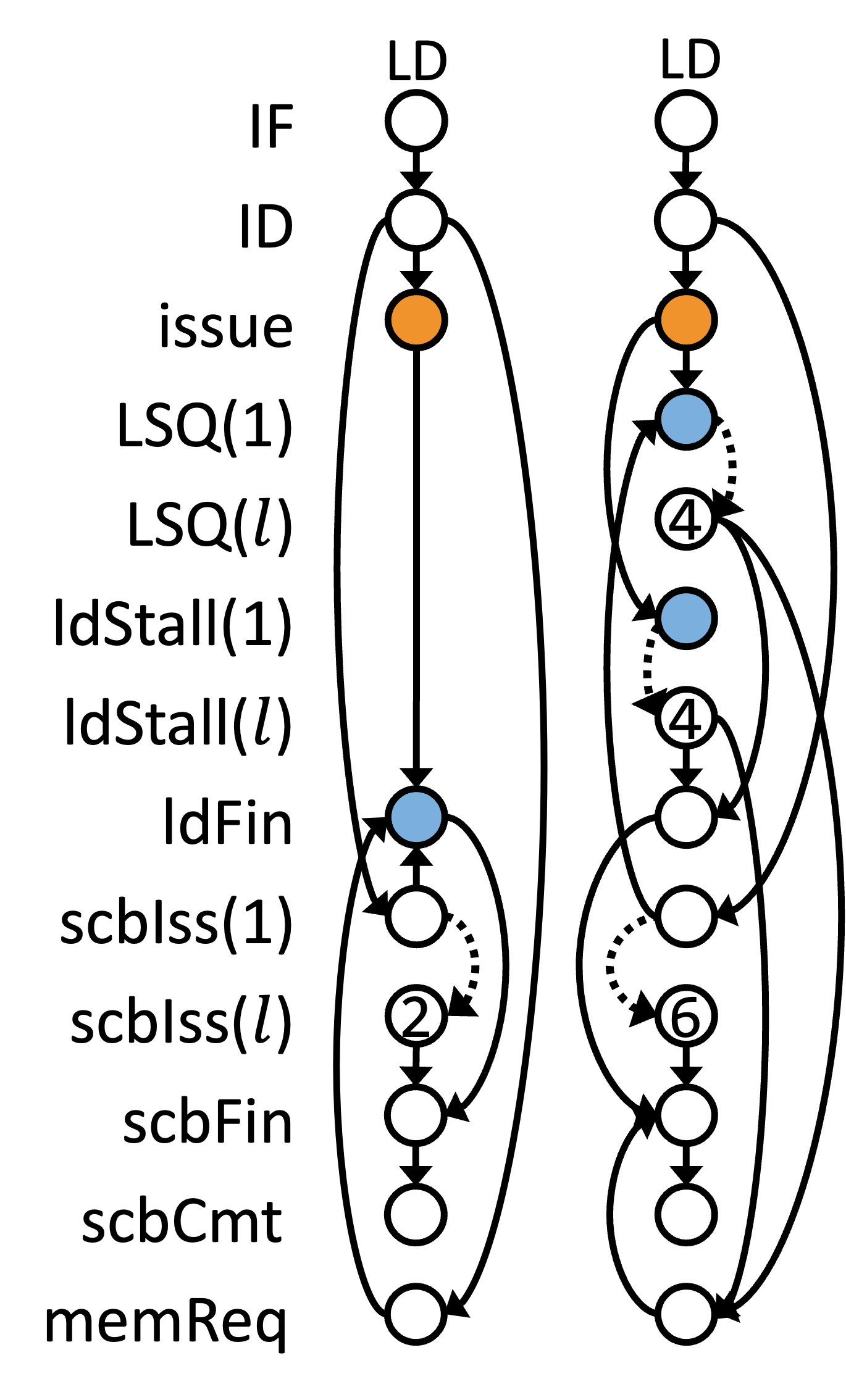}
    \caption{\inst{LD} \upaths{}}
     \label{fig:ld-upaths}
    \end{subfigure}
    \hfill
    \begin{subfigure}[t]{0.32\linewidth}
    \centering
    \includegraphics[height=5.6cm]{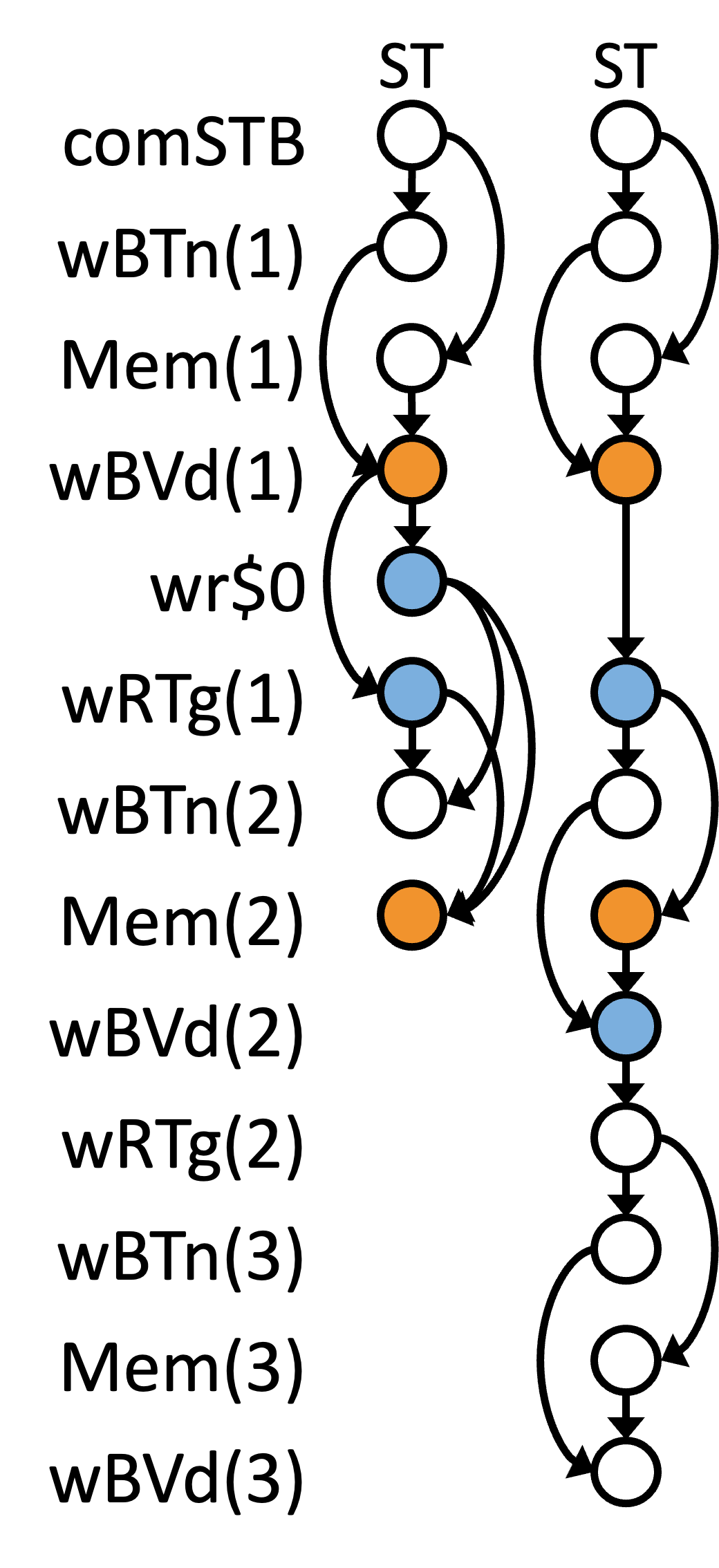}
    \caption{\inst{ST} \upaths{}}
     \label{fig:st-upaths}
    \end{subfigure}
    \caption{
    A sampling of \upaths{} for the CVA6 core (for \inst{BEQ}, \inst{LD}) and data cache (for \inst{ST}), synthesized by \PathFinder{}  (\S\ref{sec:results}). 
     Row({\inst{1}/\llt{}}): 1st/\llt{}-th visit to Row.
     Node label: value of \llt{} for \upath{}.
   }
    \label{fig:upaths}
\end{figure}

%% file: fig-f-leak.tex
\begin{figure}[t!]
    \begin{tcolorbox}[left=0pt,right=0pt,top=1pt,bottom=1pt]
    \footnotesize
        \noindent\textcolor{ForestGreen}{{\RaggedLeft// \textbf{CVA6\hbox{-}OP Core} \inst{ADD} 
        (\S\ref{sec:op-packing-example}):} \inst{ADD} (i.e., \inst{ADD^N}) in \inst{ID} is issued if it is ready (the oldest in \inst{ID}) or eligible for operand packing; else, it is stalled.}\\[2pt]
    \noindent\hlt{$\mathsf{dst\;ADD\_ID(ADD^N\;i0,\;ADD^D_{\color{black}O}\;i1)}$}$\mathsf{\;:}$ \newline
    \indent$\;\;\mathsf{\textcolor{blue}{return}\;ite(((visit(i1, ID) \;\wedge\; (\forall arg \in \{}$$\hlt{$\mathsf{i0.\iparg{arg0}, i0.\iparg{arg1}, i1.\itarg{arg0}, i1.\itarg{arg1}}$}$\newline
    \indent$\;\;\;\;\mathsf{\}, msb(arg) < 32))\;\vee\;rdy(i0)),\;}$
\hlt{$\mathsf{\{scbIss, issue\}, \{ID\}}$}$\mathsf{)}$
    \\[4pt]
    \noindent\textcolor{ForestGreen}{{\RaggedLeft// \textbf{CVA6 Core} \inst{LD} (\S\ref{sec:store-to-load-stall}): \inst{LD} in \inst{issue} finishes or is stalled depending on whether its address page offset overlaps with that of a pending \inst{ST}.}}\\[2pt]
    \noindent\hlt{$\mathsf{dst\;LD\_issue(LD^N\;i0, ST^D_{\color{black}O}\;i1)}$}$\;\mathsf{:}$ \newline
        \indent$\;\;\mathsf{\textcolor{blue}{return}\;ite((}$\hlt{$\mathsf{i1.\itarg{addr}}$}$\mathsf{\;\in (}$\hltim{$\mathsf{comSTB}$}$\;\mathsf{ \cup }$ \hltim{$\mathsf{specSTB}$}$\mathsf{) \;\wedge\;}$ \newline 
        \indent$\;\;\;\;\mathsf{offset(}$\hlt{$\mathsf{i0.\iparg{addr}}$}$\mathsf{)== offset(}$\hlt{$\mathsf{i1.\itarg{addr}}$}$\mathsf{)),}$\hlt{$\mathsf{\{ldStall, LSQ\}, \{ldFin\}}$}$\mathsf{)}$
    \\[4pt]
    \textcolor{ForestGreen}{{\RaggedLeft// \textbf{CVA6 Cache} \inst{ST} (\S\ref{sec:cva6cache}): A \inst{ST} accesses one of two data banks on a hit in the 4-way set-assoc. no-write-alloc. cache. }}\\[2pt]
    \noindent\hlt{$\mathsf{dst\;ST\_wBVld(ST^N\;i0, LD^S\;i1)}$}$\mathsf{\;:}$ \newline 
        \indent$\;\;\mathsf{hit = (}\hltim{$\mathsf{cacheTag}$}\mathsf{[set(}$\hlt{$\mathsf{i1.\itarg{addr}}$}$\mathsf{)][way] == tag(}$\hlt{$\mathsf{i1.\itarg{addr}}$}$\mathsf{)\;\wedge\; set(}$\newline 
        \indent$\;\;\;\;\mathsf{}$\hlt{$\mathsf{i0.\iparg{addr}}$}$\mathsf{) == set(}$\hlt{$\mathsf{i1.\itarg{addr}}$}$\mathsf{)\;\wedge\;tag(}$\hlt{$\mathsf{i0.\iparg{addr}}$}$\mathsf{) == tag(}$\hlt{$\mathsf{i1.\itarg{addr}}$}$\mathsf{))}$ \newline
        \indent$\;\;\mathsf{\textcolor{blue}{return}\;ite(hit, }$\hlt{$\mathsf{\{wRTag, wr\$[way/2]\}, \{wRTag\}}$}$\mathsf{)}$ 
        \\ [4pt]
    \textcolor{ForestGreen}{{\RaggedLeft// \textbf{CVA6 Core} \inst{ST} (\textit{new channel}, \S\ref{sec:cva6core}): \inst{ST} at \inst{comSTB} is stalled from draining to memory if its address offset does not match a younger \inst{LD}}.} \\ [2pt] 
    \noindent\hlt{$\mathsf{dst\;ST\_comSTB(SW^N\;i0, LD^D_{\color{black}Y}\;i1)}$}$\mathsf{\;:}$ \newline 
        \indent$\;\;\mathsf{\textcolor{blue}{return}\;ite((visit(i1, issue)\;\wedge\;offset(}$\hlt{$\mathsf{i0.\iparg{addr}}$}$\mathsf{)\;==\; offset(}$\hlt{$\mathsf{i1.\itarg{addr}}$}$\mathsf{)),\;}$\newline
        \indent$\mathsf{\;\;\;\; }$\hlt{$\mathsf{\{memRq, comSTB\},\{comSTB\}\}}$}$\mathsf{)}$
    \end{tcolorbox}
%
%
    \caption{
    Leakage function examples. 
    \hltim{Implicit inputs} and \hlt{leakage} \hlt{signature} components are highlighted. $T^\mathtt{N}/T^\mathtt{D}_\mathtt{O|Y}/T^\mathtt{S}$: intrinsic / {older or younger} dynamic / static transmitters.
    PO: program order. 
    $\mathsf{msb}$: most significant bit.
    \inst{ite(c,t,f)}: $t$ if $c$ is true; else, $f$.
    }
    \label{fig:leakage_func_examples}
\end{figure}

%% file: table-hw-defense-2.tex
{
\renewcommand{\arraystretch}{1.1}
\begin{table}[t!]
\vspace{7pt}
\setlength\tabcolsep{1pt}
\centering
\footnotesize
\begin{tabular}[]{
        |>{\centering\arraybackslash} m{19mm} 
        |>{\raggedright\arraybackslash} m{42mm}
        |>{\centering\arraybackslash} m{2mm}
        |>{\centering\arraybackslash} m{2mm}
        |>{\centering\arraybackslash} m{4.5mm}
        |>{\centering\arraybackslash} m{3mm}
        |>{\centering\arraybackslash} m{3mm}
        |>{\centering\arraybackslash} m{3mm}
        |>{\centering\arraybackslash} m{2mm}|
        }
        \hline
\multirow{2}{19mm}{\centering \textbf{Defense}} & \multicolumn{1}{c|}{\multirow{2}{42mm}{\centering\textbf{Leakage Contract Components}}} & \multirow{2}{*}{\textbf{$\mu$}} & \multicolumn{6}{c|}{\textbf{Leakage Sig.}} \\ 
\cline{4-9}
& & & 
\inst{P} & \inst{src} & $T^\mathtt{N}$ & $T^\mathtt{D}$ & $T^\mathtt{S}$  & \inst{a} 
\\ 
\specialrule{.2em}{0pt}{0pt}
CT, SCT (\S\ref{sec:background:hw-sc-defenses}) \newline SpecShield~\cite{barber:specshield} \newline ConTExt~\cite{Schwarz:ConTExT} 
& Constant-time contract (\S\ref{sec:background:hw-sc-defenses}) & - & - & - & \chkm{} & \chkm{} & \chkm{} & \chkm{} \\
\specialrule{.2em}{0pt}{0pt}
\multirow{2}{*}{MI6~\cite{Bourgeat:m16}} & Contention-based dynamic channels & - & \chkm{} & \chkm{} & \chkm{} & \chkm{} & - &  - \\ 
\cline{2-9}
& Static channels & - & \chkm{} & \chkm{} & - & - & \chkm{} & - \\
\specialrule{.2em}{0pt}{0pt}
OISA~\cite{yu:oisa} & Input-dependent arithmetic units & - & - & \chkm{} & \chkm{} & - & - & \chkm{} \\
\specialrule{.2em}{0pt}{0pt}
\multirow{5}{13mm}{STT~\cite{jiyong:stt}
 SDO~\cite{jiyong:sdo}
 SPT~\cite{choudhary:2021:spt}} & Explicit channels &-& \chkm{}& \chkm{} &\chkm{} &-&-& \chkm{} \\
 \cline{2-9}
 & Implicit channels &- & \chkm{} & \chkm{} & - & \chkm{} & \chkm{}  & \chkm{} \\
 \cline{2-9}
  & Implicit branches &- & \chkm{} & - & - & \chkm{} & \chkm{}  & \chkm{} \\
 \cline{2-9}
  & Prediction-based channels & - & \chkm{} & \chkm{} & - & - & \chkm{} & \chkm{}\\
 \cline{2-9}
 & Resolution-based channels & - & \chkm{} & \chkm{} & - &  \chkm{} & - & \chkm{} \\
\specialrule{.2em}{0pt}{0pt}
{\multirow{1}{13mm}{\centering SDO~\cite{jiyong:sdo}}} & Data-oblivious variants & \chkm{} & - & - & \chkm{} & - & - & \chkm{} \\
\specialrule{.2em}{0pt}{0pt}

\multirow{6}{13mm}{Dolma~\cite{dolma}} & Variable-time micro-ops & - & - & - & \chkm{} & - & - & \chkm{}\\
\cline{2-9}
& Contention-based dynamic channels
& - & \chkm{} & \chkm{} & \chkm{} & \chkm{} & - & \chkm{} \\ 
\cline{2-9}
& Inducive micro-ops & & \chkm{} & - & - & \chkm{} & - & \chkm{}\\ 
\cline{2-9}
& Resolvent micro-ops & - & - & - & - & \chkm{} & - & \chkm{}\\ 
\cline{2-9}
& Prediction resolution points & - & \chkm{} & \chkm{}    & - & \chkm{} & - & \chkm{}\\ 
\cline{2-9}
& Persistent state modifying micro-ops & - & - & -  & - & - & \chkm{}& \chkm{} \\ 
\hline
\end{tabular}
\caption{Six leakage contracts (\S\ref{sec:background:hw-sc-defenses}) mapped onto \upaths{} ($\mu$) and leakage signatures. \inst{a}: Arguments. \chkm{}/-:  Relevant/irrelevant to the leakage contract component.
%
%
} 
\label{table:defenses_map}
\end{table}
}

%% file: 05-tsynth-tool.tex
\section{\texorpdfstring{\synthlc{}}{SynthLC} Tool: Using \texorpdfstring{\PathFinder{}}{RTL2MuPATH} to Synthesize Leakage Signatures from Processor RTL} 
\label{sec:both-tool}

We present \synthlc{}, an automated approach and tool for synthesizing a comprehensive set of formally verified leakage signatures from a SystemVerilog processor design, from which
one can derive the leakage contracts in Table~\ref{table:defenses_map}.
First, \synthlc{} uses \PathFinder{} to uncover all \upaths{} for each instruction implemented on the design (\S\ref{sec:tsynth-tool:path-exploration}). Instructions with more than one \upath{} are \textit{candidate transponders}.
Second, \synthlc{} uses a symbolic information flow analysis to classify each candidate transponder's decisions as dependent on the unsafe operand(s) of (typed) transmitter(s) or not (\S\ref{sec:tsynth-tool:f-leak-syn}). The result is a set of true transponders with leakage signatures that characterize all of their transmitter operand-dependent \upath{} variability.

\input{fig-overview}

\subsection{Inputs \& Metadata Requirement} 
\label{sec:tool:metadata}

Both \synthlc{} and \PathFinder{} require three inputs: 
the SystemVerilog \textit{design under verification} (DUV), a list of \textit{encodings} for each implemented instruction, and design \textit{metadata}. We detail our metadata requirement below, which follows from the fact that 
these tools make extensive use of model checkers to evaluate auto-generated LTL properties, formulated as SystemVerilog Assertions (SVAs)~\cite{ieee1800}.
Table~\ref{table:annotation} quantifies this metadata for the CVA6 Core and Cache (\S\ref{sec:case-study}).


First, the designer must identify the \textit{instruction fetch register} (IFR)~\cite{rtl2uspec:hsiao:2021}, which
holds fetched instruction encodings before they are supplied by the processor frontend to the backend. 
\PathFinder{} uses the IFR to constrain the execution traces considered by a model checker,
e.g., to those that feature some specific \textit{instruction under verification} (IUV)~\cite{rtl2uspec:hsiao:2021, reid:isaformal, deutschmann:upec-do}.



\PathFinder{} uses a single IID  to track an IUV as it progresses through various PLs during its execution on the DUV (\S\ref{sec:pls})---its PC.
Hence, \PathFinder{} requires each \ufsm{}'s IIR to be a
\textit{program counter register} (PCR)~\cite{reid:isaformal,rtl2uspec:hsiao:2021,deutschmann:upec-do}, which contains the PC of the instruction occupying it.
%
%
PCRs may be present in the original DUV or added 
in parallel to the IIRs in the DUV
solely for verification ~\cite{reid:isaformal,deutschmann:upec-do,rtl2uspec:hsiao:2021}.\footnote{Such auxiliary state elements exist exclusively within the verification environment, and are removed prior to synthesis and fabrication.} 
Fig.~\ref{fig:iir_pcr} shows an excerpt of two CVA6 design files where we add PCRs in parallel to existing IIRs.


Once the DUV is augmented (if necessary) with PCRs, 
\PathFinder{} requires the user to supply all \ufsms{} as tuples of signal names $\langle \mathtt{pcr}, \mathtt{vars} \rangle$ that denote their PCR and state variable components (\S\ref{sec:pls}).
Plus, since \PathFinder{} considers invalid any PL $\langle \mathtt{\mu fsm}, \mathtt{state} \rangle$ where $\mathtt{state} = \mathit{idle}$, the user is also required to supply each $\mu \mathtt{fsm}$'s \textit{idle} state(s).
For CVA6, there are 21 PCRs, and thus \ufsm{}s, in total; we add 14 PCRs to the baseline design.

To detect when instructions commit, \PathFinder{} requires the user to supply the DUV's \inst{commit} signal.

Two final inputs support \synthlc{}'s symbolic information flow analysis. First, the user must identify 
the \textit{architectural register file} {\color{black}(ARF)} and \textit{architectural main memory} (AMEM) to block taint propagation between instruction outputs/inputs.
Second, they must identify \textit{operand registers}, located at the issue or register read stage, 
to enable taint introduction for transmitter operands.

\subsection{\texorpdfstring{\PathFinder{}}{RTL2MuPATH} Tool: Synthesizing \texorpdfstring{\upaths{}}{upaths} from RTL}
\label{sec:tsynth-tool:path-exploration}


For each IUV, taken from the input list of instruction encodings, \PathFinder{} finds all of its \upaths{} by using model checkers to explore its 
execution behavior in \textit{all reachable contexts}, {\color{black}starting from a \textit{valid reset state}, i.e.,}
a hard processor reset,  where \textit{\color{black}only} 
 architectural state is symbolically initialized. 
 All reachable contexts indicates that the IUV may be preceded/followed by an arbitrary number of valid instructions. 

\input{fig-table-annotation}

We explain \PathFinder{}'s synthesis procedure below, as depicted in Fig.~\ref{fig:overview}. Each step involves instantiating numerous SVA properties from templates. Important SVA syntax for understanding our templates includes the notions of 
\svaterm{cover} 
and \svaterm{assume} statements.
A \svaterm{cover} property directs a model checker to search for \textit{any} execution trace that satisfies a given condition. 
A \textit{reachable} outcome is returned when a trace is found. An \textit{unreachable} outcome is a proof that no such trace exists. An \textit{undetermined} outcome indicates that a satisfying trace cannot be found due to a timeout or resource constraints, but it \textit{may} exist. We discuss implications of undetermined model checker outcomes in \S\ref{sec:soundness-completeness}.
SVA \svaterm{assume} statements constrain the execution traces considered by a model checker to those that satisfy their specified condition when evaluating SVA properties (e.g., \svaterm{cover} properties).  

\subsubsection{PL Reachability for DUV}
\label{sec:tsynthtool:duvpl}
For each $\mathtt{\mu fsm} = \langle \mathtt{pcr}, \mathtt{vars} \rangle$
(\S\ref{sec:pls}), \PathFinder{} enumerates its
feasible PLs by considering all constant valuations of $\mathtt{vars}$ and excluding user-identified \textit{idle} states.
Then \PathFinder{} instantiates SVA
properties to prune 
those PLs that are proven \textit{unreachable} on the DUV \textit{by any instruction}. The remaining \textbf{DUV PLs}
are those PLs reachable by \textit{some} IUV (i.e., \textit{some} IUV can visit them). 
Our CVA6 case study has a total of 41 DUV PLs.  

Next, \PathFinder{} conducts several \textit{IUV-specific} analyses.





\subsubsection{PL Reachability for IUV}
Like the first step, but conditioned on a specific IUV, \PathFinder{} instantiates SVA properties 
to discard from consideration PLs that are proven
\textit{unreachable} by the IUV, producing a set of \textbf{IUV PLs}.
For example, in Fig.~\ref{fig:overview}, the \inst{LSQ} PL is part of the set of DUV PLs, but not included in the set of IUV PLs for an \inst{ADD}.

\subsubsection{Fine-Grained Pruning}
\label{sec:tsynth:fine-grained-pruning}
For an IUV, our first goal is to derive its \textbf{Reachable PL Sets}. A Reachable PL Set is a set of PLs that is \textit{exclusively} 
visited in one of the IUV's executions, i.e., there exists an execution of the IUV that visits all of the PLs in the set and no others. 

Deriving all {Reachable PL Sets} for an IUV, naïvely,
requires asking a model checker to consider each element in the \textit{power set} of the {IUV PLs} and deduce its reachability.
\PathFinder{} prunes this power set in two ways using SVAs: it removes elements from the power set that
contradict \textit{dominates} and \textit{exclusive} relationships between PLs.
We say $pl_0$ \textit{dominates} $pl_1$ iff all executions of the IUV that visit $pl_0$ also visit $pl_1$. 
We say $pl_0$ and $pl_1$ are (mutually) \textit{exclusive} iff there exists no execution of the IUV that visits both $pl_0$ and $pl_1$.

\PathFinder{} deduces dominates (exclusive) relationships by instantiating and evaluating the top (bottom) SVA property template  below 
for each ordered (unordered) pair of IUV PLs.
In all such listings in the paper, 
{\color{blue}\myttt{blue}} (\svaterm{brown}) 
terms are template arguments (SystemVerilog or SVA keywords). 


\noindent
\begin{minipage}{\linewidth}
\begin{lstlisting}[style=svaListing]
pl_0_dom_pl_1: cover (!^\svaarg{pl\_0}^_visited & ^\svaarg{pl\_1}^_visited);
pl_0_excl_pl_1: cover (^\svaarg{pl\_0}^_visited & ^\svaarg{pl\_1}^_visited);
\end{lstlisting}
\end{minipage}

An \textit{unreachable} outcome for the top (bottom) property proves 
there exists no execution trace where the IUV visited $pl_1$ but not $pl_0$ (visited both), thus helping to prune PL Sets.


%

\subsubsection{PL Set Reachability}
\label{sec:tsynth:iuv-pl-set-check}

For each \textbf{Candidate PL Set} that remains after pruning, \PathFinder{} instantiates this property:
%

\noindent
\begin{minipage}{\linewidth}
\begin{lstlisting}[style=svaListing]
// {pl_0, pl_1, ..., pl_n}: IUV PLs
// ^\svaarg{cand\_pl\_set}^ consists of ^\svaarg{pl\_0, pl\_1, ..., pl\_j}^
assume (!^\svaarg{pl\_\{j+1\}}^ & !^\svaarg{pl\_\{j+2\}}^ ... & !^\svaarg{pl\_n}^);
cand_pl_set: cover (^\svaarg{pl\_0}^_visited & ^\svaarg{pl\_1}^_visited & ... & ^\svaarg{pl\_j}^_visited & !(^\svaarg{pl\_0}^ | ^\svaarg{pl\_1}^ | ... | ^\svaarg{pl\_j}^));
\end{lstlisting}
\end{minipage}


A \textit{reachable} outcome indicates an execution trace exists where, by the time the IUV has disappeared from the processor (\texttt{!(\svaarg{pl\_0} | ... )}), it has visited exclusively PLs in the Candidate PL Set (\texttt{\svaarg{cand\_pl\_set}}).

Next, \PathFinder{} iterates over {\color{black}the elements of} each \textbf{Reachable PL Set} {\color{black}(discovered above)} and uses SVAs to
determine which constituent PLs may {\color{black} non-consecutively or consecutively} be revisited (\S\ref{sec:uhb-timing}).
{\color{black} Non-consecutively} revisited PLs are {\color{black}simply} marked as such.
{\color{black} Consecutively revisited PLs are duplicated and tagged as representing the first/last consecutive visit}, e.g., {\color{black}\inst{ID(1)}/}\inst{ID(}$l$\inst{)} 
in Fig.~\ref{fig:upaths}.
Knowing \textit{which} PLs may be 
{\color{black} non-consecutively or consecutively} 
revisited 
is sufficient to {\color{black} place HB edges and} uncover all \textit{decisions} across an IUV's \upaths{}---what \synthlc{} ultimately derives from \PathFinder{}'s output (\S\ref{sec:decisions}). So, \PathFinder{} {\color{black} can be configured to} avoid deducing the exact number of revisits per PL (e.g., all possible values of $l$ {\color{black}for \inst{ID(}$l$\inst{)}}) as an optimization.
{\color{black} The majority of our experiments in do this (\S\ref{sec:case-study}).}

\subsubsection{Happens-Before Edges} 
\label{sec:tsynth:hb_edge}
\PathFinder{} extends each Reachable PL Set to a full \upath{} by deriving a partial order on visited PLs. 
\PathFinder{} considers as candidate HB edges all ordered pairs of PLs that are connected via pure combinational logic in the DUV to capture  \textit{causal} happens-before relationships among visited PLs exclusively. Each candidate edge is evaluated for every Reachable PL Set using SVAs to determine if it constitutes a true HB relationship.


\subsubsection{\color{black} All Cycle-Accurate \upaths{}}
\label{sec:cycle-accurate-upaths}

{\color{black} While not needed by \synthlc{},  
\PathFinder{} can be configured to uncover: (i) 
for each IUV, for each PL that it may 
revisit, the set of \textit{revisit cycle counts} for the PL that the IUV can exhibit across all of its executions,
or (ii) all \upaths{} that concretize the precise number of visits to each revisited PL  (as in Fig.~\ref{fig:zero-skip-uhb}, ~\ref{fig:upaths}, \ref{fig:add_path_1}, and \ref{fig:add_path_2}). 
We direct \PathFinder{} to perform (i) for CVA6 to support SDO (Table~\ref{table:defenses_map}). \PathFinder{} is currently not optimized for (ii), which generates \textit{many} (still easy-to-check) properties, proportional to the cross product of distinct concrete values per PL derived in (i). We expect we can drastically prune these properties (as in \S\ref{sec:tsynth:fine-grained-pruning}), but it is unnecessary for our focus: leakage contract verification.

}

\subsection{Attributing \texorpdfstring{\upath{}}{upath} Variability to Transmitters}
\label{sec:tsynth-tool:f-leak-syn}

After \PathFinder{} uncovers all \upaths{} for each implemented instruction on the DUV, \synthlc{} identifies candidate transponders and collects all decisions for each (\S\ref{sec:decisions}).


\subsubsection{Symbolic IFT}
\label{sec:symbolic-ift}
For each candidate transponder $P$, \synthlc{} 
considers each of its decisions $(src, \mathbf{dst}) \in \mathbf{d}^P_M$ ($M$ is the DUV) in turn to determine if $P$ exhibits $(src, \mathbf{dst})$ as a function of some transmitter $T$'s unsafe operand $op$. All possible $(T, op)$ pairs are considered for each decision.

To do this, \synthlc{} first augments the DUV with cell-level \textit{information-flow tracking} (IFT) circuitry, which supports per-data-bit introduction and propagation of \textit{taint} to track the explicit and implicit flows of certain data dynamically at runtime~\cite{solt2022cellift}. Next, it uses a model checker to consider a dynamic instance of $P$, $i_P$, and a dynamic instance of $T$, $i_T$, executing together in all reachable contexts following a
{valid reset state (\S\ref{sec:tsynth-tool:path-exploration})}, 
 subject to three assumptions {\color{black}  (Fig.~\ref{fig:taint_cases})}.
In all cases, each data bit is assigned one taint bit, taint is introduced for $op$ of $i_T$ \textit{exclusively}, and taint is prohibited from propagating architecturally between instruction outputs/inputs.

\input{fig-sva-constraint}

\textit{Assumption 1} determines if $T$ is an intrinsic transmitter with respect to $(src, \mathbf{dst})$ by constraining $i_T$ and $i_P$ to be the same dynamic instruction.
\textit{Assumption 2}, which is composed of sub-assumptions 2a and 2b, determines if $T$ is a dynamic transmitter by constraining $i_T$ to be in-flight when $i_P$ visits its decision source $src$. Sub-assumption 2a/2b considers the case where $i_T$ is fetched before/after $i_P$ (i.e., $i_T$ is older/younger than $i_P$).
\textit{Assumption 3} determines if $T$ is a static transmitter by constraining $i_T$ to have materialized and dematerialized in \M{} \textit{before} $i_P$ reaches $src$.

Note that the third assumption uses one additional taint bit per data bit to 
support flushing ``sticky'' taint that is associated with $op$'s dynamic influence on transponders' \upaths{}, thereby considering its static influence exclusively.

The SVA template below has two \svaterm{assume}s. The first one introduces taint (exclusively) at the register corresponding to $op$ (\S\ref{sec:tool:metadata}), when $i_T$ is at the issue stage. The second restricts execution traces to those satisfying one of three aforementioned assumptions.
The \svaterm{cover} property searches for an execution trace where $i_P$ visits $src$ (\texttt{\svaarg{src\_pl}}) once cycle before (\texttt{\#\#1}) it visits all of the PLs in $\mathbf{dst}$ (\texttt{(\svaarg{dst\_pl\_0} \& ...)}) \textit{and} the \ufsms{} of these decision destinations are tainted---signaling a dependence on $i_T$'s operand $op$. 

\noindent
\begin{minipage}{\linewidth}
\begin{lstlisting}[style=svaListing2]
// Candidate transponder decision (*\svaarg{src\_pl}*, {*\svaarg{dst\_pl\_0}*, ...})
assume ((iT_at_issue) ^ (*\svaarg{op\_reg}*_taint == 1)); 
assume (*\svaarg{assumption 1/2/3}*); 
decision_taint: cover (*\svaarg{src\_pl}* ##1 ((*\svaarg{dst\_pl\_0}* & ...) & 
 (*\svaarg{dst\_pl\_0}*_taint | ... ))); 
\end{lstlisting}
\end{minipage}

A \textit{reachable} outcome results in assigning a \textit{tag} to $P$'s decision $(src, \mathbf{dst})$, denoting that it is dependent on typed (intrinsic / older or younger dynamic / static) transmitter $T$'s unsafe operand $op$. An \textit{unreachable} outcome assign no tag.

After using the property above to evaluate every one of $P$'s decisions (under all three assumptions, for every possible $(I, op)$ pair), we can construct a leakage signature for each of $P$'s decision sources as follows (\S\ref{sec:leakage-signatures}).
If $P$ exhibits at least two transmitter operand-dependent decisions with respect to $src \in \mathbf{src}^P_M$, we construct a leakage signature corresponding to leakage function \inst{P\_src} (function name).\footnote{A single decision may be tagged transmitter operand-dependent due to imprecision of symbolic IFT (\S\ref{sec:tsynth-tool:f-leak-syn}). At least two decisions must be operand-dependent to yield ${>}1$ receiver observations as a function of operand values.}
Examining tags assigned to all such decisions (those involving $src$) gives us typed transmitters (explicit inputs) and unsafe transmitter arguments (in the function body). Decision destinations (return values) are all sets of PLs $\mathbf{dst}$ such that $(src, \mathbf{dst}) \in \mathbf{d}^P_M$.

Notably, \synthlc{}'s symbolic IFT step also uncovers implicit inputs (Fig.~\ref{fig:leakage_func_examples}) to leakage functions, which may be useful towards implementing the hardware side-channel defenses in \S\ref{sec:background:hw-sc-defenses} using the leakage contracts in Table~\ref{table:defenses_map}.

\subsubsection{Security Argument}
\label{sec:tool:security-arg}
A proof in our repository~\cite{https://github.com/yaohsiaopid/SynthLC}
shows that \synthlc{} produces a set of leakage signatures that capture all violations of \textit{hardware side-channel safety}, as defined below, subject to a receiver \Rupath{}. \Rupath{} observes the PLs occupied by in-flight instructions in each cycle, modeling a receiver that perceives channel modulations via their impact on instruction/program execution time or resource contention.

\begin{definition}[Hardware Side-Channel Safe]
\label{def:scsafe}
A microarchitecture ${\mathnormal{M}}$ is
\textit{hardware side-channel safe}
with respect to receiver ${\mathnormal{R}}$, or
$\mathrm{SC\hbox{-}Safe}({\mathnormal{M}}, {\mathnormal{R}})$, iff:
\begin{equation}
\begin{split}
      \forall p.\forall \pi.\forall \sigma, \sigma'. \forall \mu. {ArchCtrl}(p) \implies \\
      (\sigma \approx_{\pi} \sigma'
      \implies \mathnormal{O_{R}}(\lbbar {p} \rbbar^{\langle\sigma, \mu\rangle}_{\mathnormal{M}}) = \mathnormal{O_{R}}(\lbbar {p} \rbbar^{\langle\sigma', \mu\rangle}_{\mathnormal{M}}))
\end{split}
\label{eq:non-interference}
\end{equation}
\end{definition}

Eq.~\ref{eq:non-interference} quantifies over all programs $p$ and security policies $\pi$ (which label program inputs as public or secret), all pairs of initial architectural states $\sigma, \sigma'$ and all initial microarchitectural states $\mu$.
Looking at the second line, the antecedent, $\sigma \approx_{\pi}\sigma'$, checks that the initial architectural states
are \textit{low-equivalent} with respect to $\pi$: they agree on the values of low data in $\pi$, i.e., $p$'s public data inputs.
The consequent, $\mathnormal{O_R}(\lbbar p \rbbar_{\mathnormal{M}}^{\langle\sigma, \mu\rangle}) = \mathnormal{O_R}(\lbbar p \rbbar_{\mathnormal{M}}^{\langle\sigma', \mu\rangle})$, asserts that \R{} obtains identical observation traces when $p$ runs on microarchitecture \M{} from initial states $\langle \sigma, \mu \rangle$ and $\langle \sigma', \mu \rangle$.
Given our focus on \textit{microarchitectural} (not architectural) side channels, the first line requires $p$ to feature the same sequence of instructions 
along all branches of secret-dependent control-flow instructions ($ArchCtrl(p)$). 


\input{fig-table-result}

Eq.~\ref{eq:non-interference} violations indicate that the observation trace obtained by receiver \R{} from running program $p$ with privacy policy $\pi$ on microarchitecture \M{} is \textit{indisputably} a function of $p$'s high inputs. Note that secret-dependent control-flow instructions can still behave as microarchitectural transmitters and cause Eq.~\ref{eq:non-interference} violations, e.g., if they create operand-dependent squashes. Moreover, virtually all microarchitecture, for any realistic receiver, will trigger Eq.~\ref{eq:non-interference} violations---the goal of a leakage contract is to account for them all.

%% file: fig-overview.tex
\begin{figure*}[t!]
    \centering
        \includegraphics[width=1.0\linewidth]{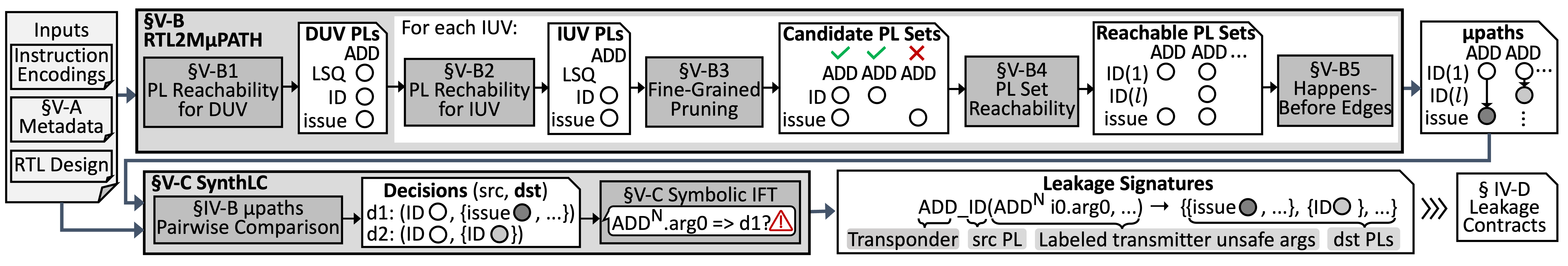}
        \caption{\PathFinder{} (top) and \synthlc{} (bottom) synthesize a complete set of formally-verified leakage signatures.
        }
        \label{fig:overview}
\end{figure*}

%% file: fig-table-annotation.tex
{
\renewcommand{\arraystretch}{1.0}
\begin{table}[t!]
\vspace{6pt}
\setlength\tabcolsep{0.8pt}
\centering
\footnotesize
\begin{tabular}[]{
 |>{\centering\arraybackslash} m{7mm}
 |>{\centering\arraybackslash} m{15mm}
 |>{\centering\arraybackslash} m{10mm}
 |>{\centering\arraybackslash} m{10.3mm}
 |>{\centering\arraybackslash} m{10mm}
 |>{\centering\arraybackslash} m{11mm}
 |>{\centering\arraybackslash} m{9mm}
 |>{\centering\arraybackslash} m{9mm}|
 }
 \hline
\multicolumn{8}{|c|}{
{\cellcolor{gray!30} \textbf{Identified in CVA6 Core}
}} \\
 \hline
 {\cellcolor{myorange!50} \inst{\textbf{IFR}}} &
 \inst{\textbf{IIRs (PCRs)} } &
 \inst{\textbf{$\mu$FSMs}} & 
 {\cellcolor{myblue!50} \inst{\textbf{PCRs}}} &
{\cellcolor{myorange!50} \inst{\textbf{commit}}} & 
{\cellcolor{myorange!50}\textbf{operand}} & 
{\cellcolor{myorange!50} \textbf{ARF}} & 
 {\cellcolor{myorange!50} \textbf{AMEM}}
 \\ \hline
 {\cellcolor{myorange!50} 1 reg} & 
 21 (7) regs & 
 38 regs & 
 {\cellcolor{myblue!50} 21 regs*} & 
 {\cellcolor{myorange!50} 1 wire} & 
 {\cellcolor{myorange!50} 2 regs} & 
  {\cellcolor{myorange!50} 1 array} & 
  {\cellcolor{myorange!50} 1 array} \\
\hline 
\end{tabular}
\\ [3pt]
\begin{tabular}[]{
 |>{\centering\arraybackslash} m{12mm}
 |>{\centering\arraybackslash} m{12mm}|
 }
\hline
\multicolumn{2}{|c|}{{\cellcolor{gray!30} \textbf{Added to Core}}} \\
\hline
\textbf{PCRs} & \textbf{SV} \\ 
\hline
14 regs & 39 LoC \\ 
\hline
\end{tabular}
%
\begin{tabular}[]{
 |>{\centering\arraybackslash} m{11mm}
 |>{\centering\arraybackslash} m{11mm}|
 }
\hline
\multicolumn{2}{|c|}{{\cellcolor{gray!30} \textbf{Added to Cache}}} \\
\hline
\textbf{PCRs} & \textbf{SV} \\ 
\hline
9 regs & 74 LoC \\ 
\hline
\end{tabular}
\begin{tabular}[]{
 |>{\centering\arraybackslash} m{15mm}
 |>{\centering\arraybackslash} m{8.5mm}
 |>{\centering\arraybackslash} m{9mm}|
 }
\hline
\multicolumn{3}{|c|}{{\cellcolor{gray!30} \textbf{Identified in CVA6 Cache}}} \\
\hline
\textbf{IIRs (PCRs)} & \textbf{PCRs} & \inst{\textbf{$\mu$FSMs}} \\ 
\hline
9 (0) regs & 9 regs* & 13 regs \\ 
\hline
\end{tabular}
%
%
 
\caption{User annotations required by \synthlc{} (\S\ref{sec:tool:metadata}) and \myhlt{myorange!50}{all}/\myhlt{myblue!50}{some}~\cite{deutschmann2023scalable, deutschmann:upec-do, dinesh2024conjunct} of prior works (\S\ref{sec:related}). *: With added ones. \ufsms{} signals correspond to their state variables (\S\ref{sec:pls}).
}
\label{table:annotation}
 \end{table}
}

%% file: fig-sva-constraint.tex
\begin{figure}[t!]
    \centering
     \includegraphics[width=0.9\linewidth]{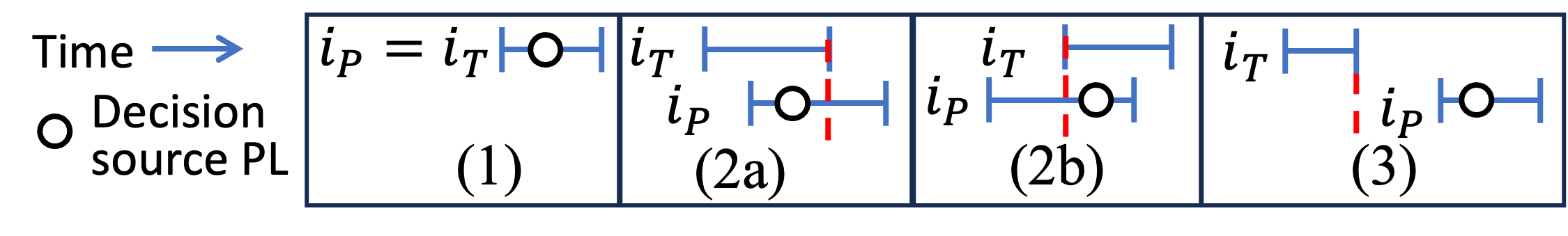}
     \vspace{-5pt}
     \caption{Constraints on SVA properties to classify transmitters as (1) intrinsic, (2a/b) older/younger dynamic, or (3) static. 
     }
     \label{fig:taint_cases}
\end{figure}

%% file: fig-table-result.tex
\begin{figure*}[!t]
        \centering
         \includegraphics[width=1.0\linewidth]{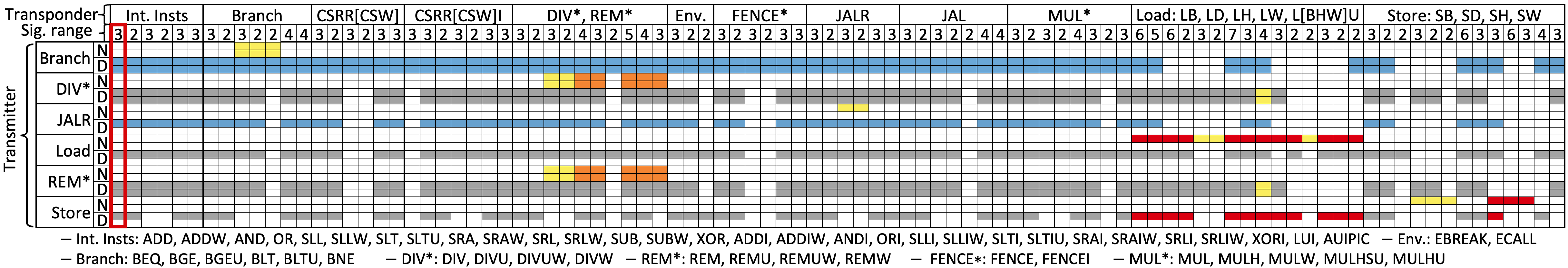}
      \vspace*{-4mm}
      \caption{
       \synthlc{}
        CVA6 Core results. Transponders 
        (coarse-grained columns) and their leakage signatures with output range sizes (fine-grained columns), plus explicit inputs 
        from intrinsic/dynamic transmitters in \inst{N}/\inst{D}-labeled rows, where the top/bottom sub-row is \inst{rs1}/\inst{rs2}. 
        We distinguish {\setlength{\fboxsep}{1pt}\colorbox{mygray}{secondary leakage\vphantom{Ay}}} (\S\ref{sec:cva6core}) and 
        {\setlength{\fboxsep}{1pt}\colorbox{myyellow}{false-positive leakage\vphantom{Ay}}} (\S\ref{sec:false-positives})
        from primary leakage (\S\ref{sec:results}). Primary leakage is categorized as involving {\setlength{\fboxsep}{1pt}\colorbox{myorange}{explicit channels}}  {\setlength{\fboxsep}{1pt}\colorbox{myred}{implicit channels}}, or 
        {\setlength{\fboxsep}{1pt}\colorbox{myblue}{explicit branches}} using  STT~\cite{jiyong:stt} terminology.
         }
       \label{fig:leakage_func_map}
\end{figure*}

%% file: 06-case-study.tex
\section{Synthesizing Leakage Signatures from CVA6}
\label{sec:case-study}




We use \synthlc{} to synthesize leakage signatures 
from the RISC-V CVA6 CPU {\color{black}(commit \#\texttt{00236BE})}, considering all 72 instructions
in the RV64I ISA and M extension (\textbf{RV64IM}). 

\label{sec:cva6}
CVA6~\cite{cva6} is a 64-bit, 6-stage, single-issue RISC-V core featuring speculation and limited out-of-order write-back with diverse functional units (ALU, LSU, Mul/Div, CSR buffer). 
%
It has a FIFO scoreboard (SCB) that tracks instructions from issue to commit and retires instructions in-order. 
Functional units may complete out-of-order in the SCB provided older, non-retired instructions’ destination registers do not match.
%

We configure the design as follows.
Both speculative and committed store buffers (STBs) are sized to two entries. 
{\color{black} The SCB is sized to four entries, but due to a bug in CVA6 that we discovered during our case study, only three entries are ever occupied at a time by active instructions (\S\ref{sec:results}).} 
Such down-scaling is typical in formal verification~\cite{dill1992protocol,dill1996downscaling}.
{We configure CVA6 without a memory management unit, 
and our main experiment instantiates the CVA6 Core as the DUV. Another experiment instantiates the CVA6 Cache (L1 data cache and cache controller) as the DUV.
}
%
CVA6  
does not come with a Verilog behavioral model of memory, so we add a single-port RAM 
behavioral model consisting of 32 64-bit words for the CVA6 Core DUV; 
the load-store unit (LSU) and committed STB are modified to directly interface with memory.


The elaboration step black-boxes the design's frontend, since \PathFinder{} drives all issued instructions at the IFR with a model checker.
The multiplier
is also black-boxed to reduce verification complexity.
Since \PathFinder{} explores control state behavior, purely combinational logic circuits can be safely abstracted---\textit{a key benefit}.
The core design features 8,577 lines of SystemVerilog; after elaboration there are 22,138 wires, 19,575 standard cells, 482 registers (11,985 D flip-flop bits), and 3 memory arrays (including ARF and AMEM in Table~\ref{table:annotation}). {The data-cache is 4-way, 128B (scaled down from 32 KB), featuring 2,279 lines of SystemVerilog.}

We supply required metadata from \S\ref{sec:tool:metadata}
to \PathFinder{} and \synthlc{} (Table~\ref{table:annotation}). 
Our implementations of both tools use SVA 2009 \cite{ieee1800} and the JasperGold v21.03 property verifier~\cite{cadence:jasper_gold}. 
We use Verific~\cite{verific} and Yosys~\cite{wolf2013yosys} to parse SystemVerilog and instrument the DUV with IFT logic using CellIFT~\cite{solt2022cellift}.
All experiments are run on three compute nodes, each of which features two 32-core 2.9GHz Intel Xeon CPUs with 512GB RAM.





%% file: 07-results.tex
\section{Results \& Discussion}
\label{sec:results}
\synthlc{} synthesizes a complete set of leakage signatures for the CVA6 Core (considering all instructions), and a partial set
for the CVA6 Cache (considering loads and stores).

\subsection{Summary of Results}
We first discuss the transmitters and transponders surfaced by \synthlc{} in our experiments.

\label{Results}


\subsubsection{Transponders and Transmitters: CVA6 Core}
\label{sec:cva6core}
Fig.~\ref{fig:leakage_func_map} summarizes \synthlc{}'s synthesis results for the CVA6 Core. Coarse-grained row/column types denote transmitters/transponders. Fine-grained row and column labels, respectively, denote transmitter types (intrinsic/dynamic) and ranges for distinct leakage signatures.
The top/bottom sub-row
for each fine-grained row indicates transmitter operand \inst{rs1}/\inst{rs2}. 

\synthlc{} finds transponders and leakage signatures per \S\ref{sec:both-tool}. 
We observe that (i) classes of transponders feature identical leakage signatures, and (ii) classes of transmitters are explicit inputs to the same leakage signatures where they feature identical types. 
So, Fig.~\ref{fig:leakage_func_map} groups transponders and transmitters accordingly.
Each fine-grained column
represents a leakage signature \inst{P\_src}, where \textit{P}
can be any transponder in the class represented by the coarse-grained column label.
Colored cells within a column 
indicate \inst{P\_src}'s explicit inputs having intrinsic/dynamic transmitters on \inst{N}/\inst{D} crossing rows. 
As an example, consider the leftmost fine-grained column, outlined in red.
It corresponds to a leakage signature \inst{ADD\_ID} that 
\synthlc{} synthesizes for \inst{ADD} transponders on the CVA6 CPU. \inst{ADD\_ID} may output one of three decisions for \inst{ADD}s with respect to decision source \inst{ID}.
The top-/bottom-most colored cell in the column indicates that
operand \inst{rs1} of a dynamic branch/store is an explicit input to  \inst{ADD\_ID}.
Overall, this column indicates that an \inst{ADD} exhibits three-way \upath{} variability
at \inst{ID} as a function of dynamic branch/division/remainder operands (both \inst{rs1} and \inst{rs2}), load/store operands (\inst{rs1}, the base address), and \inst{JALR} operands (\inst{rs1}, the target address).

Colored Fig.~\ref{fig:leakage_func_map} cells represent primary (orange, red, blue)
versus secondary (gray)  leakage.
Primary/secondary leakage indicates that the transponder (column) can/cannot leak the transmitter's (row) unsafe operand without the presence of other transponders.
%
Secondary leakage often arises due to shared resources, e.g., an \inst{ADD} that is stalled from committing at the SCB, stuck behind an intrinsic transmitter (e.g., \inst{DIV}).
\synthlc{} flags all 72 evaluated instructions as transponders and
finds that the CVA6 core features intrinsic and dynamic transmitters exclusively (hence the omission of static transmitter labels in Fig.~\ref{fig:leakage_func_map}).
\textbf{Nineteen} intrinsic transmitters are found: eight division (\inst{DIV}) and remainder (\inst{REM}) variants, seven load (\inst{LD}) variants, and four store (\inst{ST}) variants.
\textbf{Twenty-six} dynamic transmitters are found: all intrinsic transmitters plus six branch variants and \inst{JALR}.
Notably, all intrinsic transmitters except stores can exhibit execution time variability as a function of their operands. 
The paragraphs below summarize key findings, organized around classes of transponders.




\paragraph*{Load}
On CVA6, a \inst{LD} transponder may exhibit several decisions at \inst{issue}, including proceeding to destinations  \inst{\{ldFin\}}/\inst{\{LSQ, ldStall\}} as described in \S\ref{sec:store-to-load-stall} as a function of \inst{rs1} of the \inst{LD} itself (\inst{LD^N}) and \inst{rs1} of a dynamic store (\inst{ST^D_{\color{black}O}}). 
\paragraph*{Store}
A \inst{ST} transponder exhibits \upath{} variability following a PL in the committed STB (\inst{comSTB}), where it stalls if a \textit{younger} in-flight load with a different address is ready to access the single-R/W-port memory; CVA6 prioritizes serving the younger load. The leakage signature (\inst{ST\_comSTB} in Fig.~\ref{fig:leakage_func_examples}) output depends on \inst{rs1} of the \inst{ST} itself (\inst{ST^N}) and \inst{rs1} of a dynamic load (\inst{LD^D_{\color{black}Y}}).
\textit{We are the first to uncover this channel} when conducting CV6 leakage contract verification~\cite{deutschmann:upec-do,deutschmann2023scalable}.

Interestingly, this channel renders CVA6 susceptible to a 
\textit{new} class of speculative interference attacks~\cite{Behnia:speculative-interference}, involving \textit{transient dynamic transmitters} (\inst{LD}s, in the shadow of older excepting instructions) that create \upath{} variability for older, \textit{committed} transponders (\inst{ST}s).
Since this \upath{} variability takes place after \inst{ST}s commit, it does not impact their execution time. 
Classic variants~\cite{Behnia:speculative-interference} involve \textit{transient intrinsic transmitters} whose own \upath{} variability creates timing-differentiable \upath{} variability for older, \textit{bound-to-commit} transponders. Using transponders, we can more generally define speculative interference attacks as involving transient transmitters that create \upath{} variability for older non-transient transponders.

\paragraph*{Division/Remainder}
\synthlc{} flags all \inst{DIV}/\inst{REM} variants as intrinsic transmitters, and thus, transponders. 
Both use serial division circuitry, taking one to sixty-six cycles to compute their results {\color{black} (based on revisit cycle counts, \S\ref{sec:cycle-accurate-upaths}).}
%


\paragraph*{All}
All transponders (all instructions) can be stalled in \inst{ID} (from issuing) 
or \inst{scbFin} (from committing)
as a function of the operand(s) of dynamic 
\inst{LD}, \inst{ST}, \inst{DIV}, and/or \inst{REM} transmitters. 
{\color{black} The stall in \inst{ID}/\inst{scbFin}  
can be 1 to 68/4 consecutive cycles.}
They may also be flushed at almost any PL as a function of dynamic branch or \inst{JALR} transmitter operands: all six branches and \inst{JALR} are flagged as dynamic transmitters. 
{\color{black} The one exception is \inst{LD} transponders; once \inst{LD}s visit certain PLs in the load unit (\inst{ldStall}, \inst{ldFin}), they cannot be flushed until they exit the load unit.
 }
Branches, as a function their \inst{rs1} and \inst{rs2} operands, and \inst{JALR}, as function of its \inst{rs1} operand, flush a transponder 
upon a mis-prediction. 
%
Prior work classifies branch and \inst{JALR} operands as unsafe on CVA6, but cannot deduce why~\cite{deutschmann:upec-do}. 
Note that no control-flow instruction is flagged as a static transmitter, because we black-box the CVA6 Core front-end (\S\ref{sec:case-study}), where  predictor structures reside.

\subsubsection{Transponders and Transmitters: CVA6 Cache} 
\label{sec:cva6cache}
{\color{black}
\synthlc{} is \textit{not} a modular verification procedure. However, we use the \synthlc{} approach to conduct a \textit{conservative} and \textit{partial} security evaluation of the CVA6 Cache in order to show that it can: 
(i) analyze a realistic cache, which no prior leakage contract verification work has attempted~\cite{deutschmann:upec-do, wang2023specification, deutschmann2023scalable, gleissenthall:iodine, gleissenthall:xenon}; (ii) handle \textit{non-consecutive} re-visit behavior (\S\ref{sec:uhb-timing}), which exists in the Cache DUV only; (iii) benefit from modularity from a scalability perspective (\S\ref{sec:result-property}).}


{\color{black}
First, \synthlc{} collects all \inst{LD}/\inst{ST} decisions based on the \PathFinder{}'s analysis results on the Cache DUV. 
Second, we select three source PLs apiece for \inst{LD}/\inst{ST}---the three with the highest number of destination PL sets (four on average). 
Third, we instantiate a Core+Cache DUV and check that these reachable Cache decisions are reachable on the full design; all pass this check in our experiment.
Finally, \synthlc{} conducts its symbolic IFT step on the Cache DUV to produce leakage signatures for all six decision source PLs.}






{\color{black}
Interestingly, all leakage 
signatures for \inst{LD} (\inst{ST}) transponders have \textit{identical} explicit inputs, which flag every relevant transmitter type---intrinsic/dynamic/static \inst{LD}s (\inst{ST}s) and dynamic/static \inst{ST}s (\inst{LD}s)---as leaking its address operand.
Plus, \synthlc{} uncovers channels involving hardware structures in nearly all Cache files, specifically: tag banks, fully-associative write buffer, MSHRs, shared ports to the AXI interface.
}


Fig.~\ref{fig:st-upaths} shows some \upaths{} for stores. A \inst{ST} visiting \inst{wBVd}, where it accesses the cache, may exhibit several decisions, including progressing to destinations 
\inst{\{wRTg,wr\$0\}}/\inst{\{wRTg\}} in the left/right \upath{} upon a cache miss/hit.
The synthesized leakage signature \inst{ST\_wBVd}, shown 
in Fig.~\ref{fig:leakage_func_examples}, 
flags \inst{LD}s as \textit{static} transmitters (\inst{LD^S}), but not \inst{ST}s (since the cache is no-allocate on write), and the \inst{ST} itself as an intrinsic transmitter (\inst{ST^N}).

{\color{black}
Results from our Cache evaluation are conservative (sound but incomplete, \S\ref{sec:soundness-completeness}), since symbolic IFT can \textit{possibly} flag transmitter-transponder interactions that are possible on the Cache DUV, but not the Cache+Core. We manually inspect results and confirm that all flagged leakage looks plausible.
}

\subsection{Discussion}
\label{sec:discussion}
\subsubsection{False-Positives from IFT}
\label{sec:false-positives}
\synthlc{} exhibits some false positives (Fig.~\ref{fig:leakage_func_map})  due to IFT imprecision.
Interestingly, it \textit{does not} identify any false-positive transmitters. 
It \textit{does}, however, identify a handful of false-positive transmitter-transponder associations, i.e., indicating that some transponder's decision is operand-dependent on some (intrinsic/dynamic) transmitter's operand when it is not.
In particular, 14/94 (1/6) unique leakage signatures obtained from the CVA Core (Cache) include extraneous explicit inputs.
So,
a few extraneous (benign) dynamic/intrinsic transmitters are flagged. 
We find these cases arise when distinct \ufsms{} in the same structure (e.g., in different SCB  entries) share fan-in signals,
which are updated as a function of transmitter operands, causing over-taint.
\subsubsection{CVA6 Bugs}

\PathFinder{}/\synthlc{} helped us identify three new functional bugs in CVA6 (two have security implications), 
 involving \inst{JAL}/\inst{JALR}/branches~\cite{git-commits}.

\PathFinder{} finds that following its visit to the \inst{scbFin} (\textit{SCB finished} PL), \inst{JALR} never progresses to \inst{scbExcp} (\textit{SCB exception} PL), while \inst{JAL} and 
branches sometimes do. 
The RISC-V ISA requires that \inst{JAL}/\inst{JALR}/branches all trigger exceptions if their target addresses are not 4-byte aligned. 
From inspection, we find CVA6 does not enforce any alignment restrictions for \inst{JALR} (as its \upaths{} indicate). 
While investigating \inst{JALR}, we also notice that \inst{JAL} only enforces 2-byte alignment checks. 
Notably, these functional bugs can expand the attack surface for control-flow hijacking attacks~\cite{carlini2015control}.

\synthlc{} reports that whether a conditional branch 
progresses  to \inst{scbCmt} (\textit{SCB commit} PL)
or \inst{scbExcp} following \inst{scbFin} is 
\textit{independent} of its operands. But, RISC-V requires that branches raise misaligned target exceptions only when their (operand-dependent) outcome is taken. 
From inspection, 
we find that branches incorrectly raise exceptions whenever their target address is misaligned, regardless of their outcome.

From the RTL waveforms produced by \PathFinder{}'s reachable SVA cover properties (\S\ref{sec:tsynth-tool:path-exploration}), we observe that the SCB is always underutilized by one entry. We localized this counterintuitive behavior to an incorrect counter width declaration in the CVA6 Core. By the time we noticed this microarchitectural bug, it had been fixed (commit \#\texttt{5c0dc19}).


\subsubsection{Property Evaluation Performance}
\label{sec:result-property}
For \upaths{} synthesis for the CVA6 Core, \PathFinder{} evaluates 124,459 properties in 4.43 minutes per property on average under a 30 minute time-out;
16.39\% of properties per instruction are undetermined. \PathFinder{}/\synthlc{} can be configured to interpret undetermined model checker outcomes (\S\ref{sec:tsynth-tool:path-exploration}) as \textit{reachable} or \textit{unreachable}. We do the latter (\S\ref{sec:soundness-completeness}). 
For leakage signature synthesis for the Core, \synthlc{} evaluates 30,774 (additional) properties for 2.35 minutes per property on average under the same time-out; 13.74\% are undetermined.
{\color{black} Interestingly, for the 
Cache, \textit{all} 4,178 properties evaluated by \PathFinder{}/\synthlc{} 
 complete within 3 seconds on average, 
 highlighting the benefits of modularization.}
%
We use \PathFinder{} to uncover revisit cycle counts
(\S\ref{sec:cycle-accurate-upaths}) on the Core only, evaluating
8,043 additional properties in 32 seconds on average with a 10 minute time-out; 0.7\% are undetermined. 

\subsubsection{Soundness/Completeness}
\label{sec:soundness-completeness}
{\color{black} 
When discussing \textit{theoretical} guarantees of \PathFinder{}/\synthlc{}, we assume that there are no undetermined model checker outcomes. Recall that all SVAs are evaluated from a valid reset state of the DUV (\S\ref{sec:tsynth-tool:path-exploration}).

The \PathFinder{} procedure is theoretically sound: if it outputs \upath{} \textit{p} for instruction \textit{I}, then an execution trace
where \textit{I} exhibits \textit{p}, was deemed \textit{reachable} on the DUV. 
And it is theoretically complete: if it does not output \upath{} \textit{p} for instruction \textit{I}, then such a trace 
was deemed \textit{unreachable}. 

Interpreting undetermined model checker outcomes as unreachable (\S\ref{sec:result-property}) impacts \PathFinder{}’s completeness guarantee. If \PathFinder{} does not output \upath{} \textit{p} for instruction \textit{I} due to an undetermined outcome, there is a chance that \textit{I} exhibits \textit{p} in some reachable trace.
Our manual inspection of \PathFinder{}’s output for CVA6 suggests most  undetermined \upaths{} would eventually resolve as unreachable.
Such \upath{}s usually feature instructions visiting PLs in unrelated functional units (e.g., an \inst{ADD} visiting a STB PL).

The \synthlc{} procedure is theoretically sound:
If it classifies some candidate transponder \textit{P}'s decisions at source PL $src$ as independent of some transmitter \textit{T}’s unsafe operand \textit{op}, then it is indeed independent. That is, an execution trace (on the IFT-augmented DUV), 
where $op$ introduces taint, \textit{P} exhibits decision $(src, \mathbf{dst})$, and $\mathbf{dst}$ is becomes tainted, was deemed \textit{reachable} for \textit{at most one} such decision and \textit{unreachable} for all others (\S\ref{sec:symbolic-ift}).
However, \synthlc{} is theoretically incomplete: such an execution trace may be deemed \textit{reachable} for \textit{more than one} such decision even if \textit{P}'s decisions at $src$
are \textit{independent} of $op$, 
due to imprecision of symbolic IFT.

We are optimistic that we can leverage established IFT techniques to minimize false positive leakage flagged by \synthlc{}, e.g., custom taint propagation rules to reduce taint spread~\cite{schwartz2010all} or specialized taint flushing mechanisms to prevent over-taint in shared buffers (e.g., the SCB)~\cite{slowinska2009pointless}.

Interpreting undetermined model checker outcomes as unreachable 
impacts \synthlc{}’s soundness guarantees.
That is, if \synthlc{} classifies some transponder’s decision at some source PL as independent of some transmitter’s unsafe operand due to an undetermined outcome, it may actually be dependent in some reachable trace.}
Nevertheless, our cache evaluation suggests that this issue can be addressed through DUV decomposition and modular verification. 

In practice, \synthlc{} automatically localizes side-channel leakage in RTL {\color{black} with exceptional precision.}

%% file: 08-related.tex
\section{Related Work}
\label{sec:related}

\paragraph*{Contract Verification}


No existing approaches can verify hardware adherence to leakage contracts as detailed as leakage signatures. The closest prior work verifies hardware adherence to CT contracts~\cite{fadiheh:upec, deutschmann:upec-do, wang2023specification, dinesh2024conjunct} by
%
checking some version of Eq.~\ref{eq:non-interference} \textit{in one shot} with varying receiver assumptions using a \textit{product circuit} formulation---\textit{two copies} of the DUV are instantiated and analyzed together.
These methodologies are all {\color{black}theoretically} incomplete due to their use of {\color{black} \textit{symbolic initial states}}
(\textit{all} registers are symbolic), which enable the product circuit approach to scale at the cost of \textit{false counterexamples}. Notably, the user must manually inspect counterexamples  to classify them as false and add constraints to avoid them.
All but UPEC-DIT~\cite{deutschmann:upec-do} are {\color{black} theoretically} sound, i.e., ignoring model-checker limitations.

{
\color{black}
UPEC-DIT~\cite{deutschmann:upec-do}
is a
\ct{} \textit{contract synthesis} approach. 
}
\textsc{LeaVe}~\cite{wang2023specification} and UPEC-DIT-23~\cite{deutschmann2023scalable} are {\color{black}CT} \textit{contract satisfaction} approaches.
{\color{black}\textsc{ConjunCT}~\cite{dinesh2024conjunct} is both.}

UPEC-DIT~\cite{deutschmann:upec-do} inputs include the DUV's operand registers, \inst{commit} signal, {\color{black}existing} PCRs, ARF, AMEM, and IFR (\S\ref{sec:tool:metadata}). 
During synthesis, the user manually classifies registers as \textit{control}/\textit{data} upon counterexample inspection, attributes control value divergence to instruction operands, and writes assumptions to avoid spurious counterexamples.
For CVA6, this amounts to 
refining over a hundred lines of SVA properties~\cite{upec_dit_repo}.


\textsc{LeaVe}'s inputs include all of UPEC-DIT's, except PCRs, plus design invariants and extra internal design signals that may hold unsafe transmitter operands.
It tries to automatically refine the invariants to construct an inductive proof, but requires updating them upon a (manually detected) spurious proof failure. 
Note, ``[\textsc{LeaVe}] does not yet scale to processors of the complexity of, e.g., the CVA6 core''~\cite{mohr2024synthesizing}. 
Over twenty lines of invariants are needed for an Ibex core variant~\cite{leave_repo,ibex}.




UPEC-DIT-23's inputs {\color{black} include UPEC-DIT's plus a list of transmitters}. 
During verification, the user writes
invariants to exclude false counterexamples and performs manual tasks from UPEC-DIT. 
{\color{black}Over a hundred of lines of invariants~\cite{upec_dit_arxiv_repo} are needed in addition to the property refinement from UPEC-DIT.}

{\color{black}
\textsc{ConjunCT}~\cite{dinesh2024conjunct} inputs are the same as UPEC-DIT's.
Its synthesis step 
classifies each instruction as  
a candidate (non-)transmitter via a bounded analysis.
Its verification
step uses automatically-generated candidate invariants to conduct unbounded verification that the set of non-transmitters indeed contains no transmitters.
Upon misclassification (during synthesis) or proof failure (during verification), the user inspects traces and constructs assumptions to avoid false positive transmitters or invariants to derive complete proofs, respectively.
The authors evaluate in-order pipelines, where few non-transmitters are misclassified as transmitters.
One pipeline required carefully-constructed assumptions to cull false positives due to unreachable states. 
Out-of-order designs (that use SCBs, ROBs, etc.) will require more of these constraints, and therefore commensurately more manual effort.
}

%% file: 09-conclusions.tex
\section{Conclusion}
\label{sec:conclusion}
We design the first automated approach and tool for uncovering all microarchitectural execution paths for instructions as implemented on a particular SystemVerilog processor design. In doing so, we observe that such path variability can be used to localize hardware side channels---a requirement for designing defenses against hardware side-channel attacks. Subsequently, we design the first automated approach and tool for synthesizing microarchitectural leakage contracts, required by ten defenses, from processor RTL.

%% file: 10-ack.tex
\section*{Acknowledgements}
We thank Mohammad Rahmani Fadiheh, Sally Wang and the anonymous reviewers for their constructive comments and feedback. This work was supported in part by the National Science Foundation (NSF), under awards 2153936, 
1954521,
1942888,
2154183,
8191902, 2321489,
and 2236855 (CAREER), and the Defense Advanced Research Projects Agency (DARPA) under contract W912CG-23-C-0025 and subcontract from Galois, Inc. 
We gratefully acknowledge a Verific Design Automation academic license and gifts from Apple and Intel.

%% file: 11-ae.tex
\section{Artifact Appendix}
{

\subsection{Abstract}

This artifact uses \PathFinder{} and \synthlc{} to conduct multi-\upath{} and leakage signature synthesis, respectively, on the RISC-V CVA6 CPU~\cite{cva6}.

\subsection{Artifact check-list (meta-information)}


\small
\begin{itemize}
  \item {\bf Data set: }
    \begin{itemize}
        \item Original CVA6 SystemVerilog design as the \textit{design under verification} (DUV)
        \item IFT-instrumented CVA6 SystemVerilog design as the DUV
        \item \textit{\PathFinder{} and \synthlc{}} code base, implemented using Python3, SVA, and TCL.
    \end{itemize}
    
    
  \item {\bf Run-time environment: }
  \begin{itemize}
      \item Cadence JasperGold: For evaluation of \textit{SystemVerilog Assertion} (SVA) properties generated by both tools.
 
  \end{itemize}

  \item {\bf Output: }
    \begin{itemize}
        \item Complete runs of \PathFinder{} and \synthlc{} on CVA6 with respect to a subset of instructions from the RISC-V ISA (\inst{ADD, DIV, LW, SW, BEQ}) to derive the following: 
        1) \upaths{} (Fig.~\ref{fig:add_path_1}, \ref{fig:add_path_2}, and \ref{fig:upaths}) for these instructions, and 2) leakage signatures corresponding to this subset of the ISA. 
        \item Complete run of \synthlc{} seeded with \upaths{} for the whole ISA to reproduce the leakage signatures as in Fig.~\ref{fig:leakage_func_map}.  
        
    \end{itemize}
  \item {\bf Approximate total time: } 
  The execution time primarily depends on how many JasperGold jobs the machine can run in parallel, which depends on the core number and the memory of the machine. 
  The execution time provided below is tested on a machine with 128 cores and 700GB memory and configured to run $N=3$ jobs in parallel. 
  For a machine with only 48-64 cores, we recommend configuring the machine to run $N=2$ jobs in parallel (more details in \S\ref{appendix:flow}). 
  Lastly, the execution time reported below roughly scales with a factor of $3/N$. 
  \begin{itemize}
        \item For the first two experiments (\upaths{} and leakage signatures for \inst{ADD, DIV, LW, SW, BEQ}), $\sim$47 ($\sim$71) hours if the machine is configured to run $N=3(2)$ JasperGold jobs in parallel. 
        \item For the third experiment (complete reproduction of all leakage signature in Fig.~\ref{fig:leakage_func_map}), the total time can take over 16 days. 
        But this step is incremental and can produce results at a rate of about one leakage signature (a column of the Fig.~\ref{fig:leakage_func_map}) every 10$\times 3/N$ $\sim$ 40$\times 3/N$ hours when machine runs N jobs in parallel, depending on the number of decisions the leakage signatures control. 
        While one can stop early, to produce minimal set of results as discussed in our experiment workflow (\S\ref{appendix:flow}) will take minimally 100 hours or so. 
    \end{itemize}
    In summary, \textbf{the total runtime is around minimally 150 hours} to see the results discussed in the instructions files (\S\ref{appendix:flow}).
  \item {\bf Archived (provide DOI)?: }  \\ 
     \url{https://doi.org/10.5281/zenodo.13288445}
\end{itemize}

\subsection{Description - How to access}

All files including data set, code base, and instructions can be found at 
\url{https://github.com/yaohsiaopid/synthlc}. 

\subsection{Software Dependencies}
\begin{itemize}
    \item JasperGold for SVA evaluation
    \item Python3 and packages including networkx, cvc5, pandas, and matplotlib for the execution of \PathFinder{} and \synthlc{}
    \item Graphviz for visualization
\end{itemize}

\subsection{Installation}
\begin{itemize}
    \item Please follow the steps in this file to install and check software dependencies: \href{https://github.com/yaohsiaopid/synthlc/blob/master/00-installation.md}{\texttt{00-installation.md}} 
\end{itemize}
\subsection{Experiment workflow}
\label{appendix:flow}
The cloned repository includes a series of instruction files at the top level.
They will walk you through the evaluation of \PathFinder{} and \synthlc{} on CVA6. Please follow these instructions in the following order:
\href{https://github.com/yaohsiaopid/synthlc/blob/master/01-setup.md}{\texttt{01-setup.md}},
\href{https://github.com/yaohsiaopid/synthlc/blob/master/02-duvpl-dfg.md}{\texttt{02-duvpl-dfg.md}},
\href{https://github.com/yaohsiaopid/synthlc/blob/master/03-rtl2mupath.md}{\texttt{03-rtl2mupath}}, 
\href{https://github.com/yaohsiaopid/synthlc/blob/master/04-synthlc.md}{\texttt{04-synthlc.md}}, 
\href{https://github.com/yaohsiaopid/synthlc/blob/master/05-5instn-isa.md}{\texttt{05-5instn-isa.md}}, and
\href{https://github.com/yaohsiaopid/synthlc/blob/master/06-lc-table.md}{\texttt{06-lc-table.md}}. 


\subsection{Evaluation and expected results}
\label{sec:instructions}
\begin{enumerate}
    \item \href{https://github.com/yaohsiaopid/synthlc/blob/master/01-setup.md}{\texttt{01-setup.md}}: 
    This instruction file walks through annotation preparation for CVA6 (\S\ref{sec:tool:metadata}), design augmentation (Table~\ref{table:annotation}), and formal environment setup for SVA property evaluation. \\ 
    We also provide instruction to configure $N$, the number of jobs the machine will run in parallel during SVA property evaluation steps. The estimated runtime mentioned below assumes $N=3$.  
    \item \href{https://github.com/yaohsiaopid/synthlc/blob/master/02-duvpl-dfg.md}{\texttt{02-duvpl-dfg.md}}: 
    This instruction file walks through DUV PL derivation (\S\ref{sec:tsynthtool:duvpl}) and DFG Analysis (\S\ref{sec:tsynth:hb_edge}), both of which are used by \PathFinder{} to explore the execution behavior of all IUVs. 
    \item \href{https://github.com/yaohsiaopid/synthlc/blob/master/03-rtl2mupath.md}{\texttt{\{03-rtl2mupath,}}\href{https://github.com/yaohsiaopid/synthlc/blob/master/04-synthlc.md}{\texttt{04-synthlc\}.md}}: 
    These two instruction files compose the first experiment of this artifact. They explain how to run the \textit{end-to-end} flow of \PathFinder{} and \synthlc{} on a \inst{DIV} instruction under a restricted execution assumption, which enables the entire experiment finish within seven hours (without this assumption, this may take many days). 
    \\
    Specifically,  \PathFinder{} is configured to explore the execution of \inst{DIV} when it is issued at the first cycle after the valid reset state (\S\ref{sec:tsynth-tool:path-exploration}) and followed by no other valid instructions; while \synthlc{} takes in these restricted set of \upaths{}, but assumes \inst{DIV} can be preceded/followed by any of the five instructions mentioned earlier (\inst{ADD, DIV, LW, SW, BEQ}) to synthesize a set of leakage signatures.  \\ 
    This experiment illustrates the detailed flow of Fig.~\ref{fig:overview}, showcasing properties automatically generated and evaluated (\S\ref{sec:tsynth-tool:path-exploration} and \S\ref{sec:synth-lc-approach}). 
    It reproduces the following key results (\S\ref{sec:results}):
    \begin{itemize}
        \item \PathFinder{} automatically uncovers sixty-six cycle accurate \upaths{} for \inst{DIV}, a subset of all \upaths{} uncovered in our full case study (\S\ref{sec:case-study}), but a sufficient amount to supply to and demonstrate the functionality of \synthlc{}. 
        \item \synthlc{} synthesizes two leakage signatures from this restricted set of \inst{DIV} \upaths{}
        and labels \inst{DIV} as an intrinsic and dynamic transmitter.
        \item \synthlc{} finds \inst{DIV} is transponder for \inst{BEQ} and \inst{LW/SW} dynamic transmitters as a function of their \inst{rs1/rs2} and \inst{rs1} operands, respectively. 
    \end{itemize}
    \item \href{https://github.com/yaohsiaopid/synthlc/blob/master/05-5instn-isa.md}{\texttt{05-5instn-isa.md}}: The second experiment of this artifact steps through a  reproduction of the \upaths{} in Fig.~\ref{fig:add_path_1}, \ref{fig:add_path_2}, and \ref{fig:upaths}. 
    To shorten runtimes, 
    we consider a restricted RISC-V ISA 
    that implements the following five instructions: \inst{ADD}, \inst{BEQ}, \inst{LW}, \inst{SW}, and \inst{DIV}. 
    This experiment can take a total of 40 hours. 
    \item \href{https://github.com/yaohsiaopid/synthlc/blob/master/06-lc-table.md}{\texttt{06-lc-table.md}}:
    \label{sec:lc-table}
    The last experiment of this artifact reproduces Fig.~\ref{fig:leakage_func_map}. First, we include in our dataset the full set of \upaths{} synthesized by \PathFinder{} for  CVA6 (for all 72 instructions in the RV64IM ISA) in our submission-time case study (\S\ref{sec:case-study}), since \PathFinder{} can take a significant amount of time to explore all 72 instructions.
    Second, given these \upaths{}, this experiment will deploy \synthlc{} to incrementally synthesize a comprehensive set of leakage signatures (Fig.~\ref{fig:leakage_func_map} columns) one at a time using SVA property generation and evaluation (\S\ref{sec:symbolic-ift}). \textit{The flow can take over a week or more to finish depending on one's machine.} One can stop the process early to observe a partial version of Fig.~\ref{fig:leakage_func_map}.
    This experiment primarily aims to showcase the details of \synthlc{}, which is the culminating contribution in this paper. 
    
\end{enumerate}





}